\documentclass[letterpaper,10pt,journal,twoside]{IEEEtran}
\IEEEoverridecommandlockouts
\usepackage{amsmath,amssymb,amsfonts}
\usepackage{algorithmic}
\usepackage[linesnumbered,ruled]{algorithm2e}
\usepackage{graphicx}
\usepackage{textcomp}
\usepackage{xcolor}
\usepackage{threeparttable}
\usepackage{tabularx}
\usepackage{booktabs}
\usepackage{multirow}
\usepackage{colortbl}
\usepackage{enumitem}
\usepackage{subfigure}
\usepackage{pifont}
\usepackage{comment}
\usepackage{array} 

\usepackage{bigstrut,multirow,rotating}
\usepackage[numbers,sort&compress]{natbib}
\usepackage{tikz}
\usepackage{threeparttable}
\usepackage[colorlinks, linkcolor=magenta, anchorcolor=green, citecolor=blue]{hyperref}

\newcommand{\hm}{\textcolor{magenta}}
\usepackage[normalem]{ulem}   
\usepackage{makecell}

\def\BibTeX{{\rm B\kern-.05em{\sc i\kern-.025em b}\kern-.08em
    T\kern-.1667em\lower.7ex\hbox{E}\kern-.125emX}}
    
\newcommand*{\circled}[1]{\lower.7ex\hbox{\tikz\draw (0pt, 0pt)%
    circle (.5em) node {\makebox[1em][c]{\small #1}};}}

\begin{document}

\title{\huge

Efficient SRAM-PIM Co-design by Joint Exploration of Value-Level and Bit-Level Sparsity

}

\author{Cenlin~Duan,
        Jianlei~Yang,~\IEEEmembership{Senior Member,~IEEE,}
        Yikun~Wang,
        Yiou~Wang,
        Yingjie~Qi,
        Xiaolin~He,
        Bonan~Yan,~\IEEEmembership{Member,~IEEE,}
        Xueyan~Wang,~\IEEEmembership{Member,~IEEE,}
        Xiaotao~Jia,~\IEEEmembership{Member,~IEEE,}
        and~Weisheng~Zhao,~\IEEEmembership{Fellow,~IEEE}
\thanks{Manuscript received on January 2025, revised on March and May 2025, and accepted on May 2025. This work is supported in part by the National Natural Science Foundation of China (Grant No. 62072019, No. 92364201, and No. 62474016), the Beijing Natural Science Foundation (Grant No. L243031), the National Key R\&D Program of China (Grant No. 2023YFB4503704 and 2024YFB4505601), and the Young Elite Scientists Sponsorship Program by CAST (2023QNRC001). 
\textit{Corresponding authors are Jianlei Yang and Weisheng Zhao.}
}
\thanks{C. Duan, X. Wang, X. Jia and W. Zhao are with Fert Beijing Research Institute, School of Integrated Circuit Science and Engineering, Beihang University, Beijing, 100191, China. E-mail: \url{weisheng.zhao@buaa.edu.cn}}
\thanks{J. Yang, Yikun Wang, Yiou Wang, Y. Qi and X. He are with School of Computer Science and Engineering, Beihang University, Beijing 100191, China, and Qingdao Research Institute, Beihang University, Qingdao 266104, China. E-mail: \url{jianlei@buaa.edu.cn}}
\thanks{B. Yan is with Institute for Artificial Intelligence, Peking University, Beijing, 100871, China.}
}

\maketitle


\begin{abstract}

Processing-in-memory (PIM) architectures mitigate the Von Neumann bottleneck by integrating computation units into memory arrays.
Among PIM architectures, digital SRAM-PIM has become a prominent approach, directly integrating digital logic within the SRAM array.
However, the rigid crossbar architecture and full array activation pose challenges in efficiently utilizing value-level sparsity.
Moreover, neural network models exhibit a high proportion of zero bits within non-zero values, which remain underutilized due to architectural constraints.
To overcome these limitations, we present Dyadic Block PIM (DB-PIM), a groundbreaking algorithm-architecture co-design framework to harness both value-level and bit-level sparsity. 
At the algorithm level, our hybrid-grained pruning technique, combined with a novel sparsity pattern, enables effective sparsity management. 
Architecturally, DB-PIM incorporates a sparse network and customized digital SRAM-PIM macros, including input pre-processing unit (IPU), dyadic block multiply units (DBMUs), and Canonical Signed Digit (CSD)-based adder trees.
It circumvents structured zero values in weights and bypasses unstructured zero bits within non-zero weights and block-wise all-zero bit columns in input features.
As a result, the DB-PIM framework skips a majority of unnecessary computations, thereby driving significant gains in computational efficiency.
Experimental results demonstrate that our DB-PIM framework achieves up to $8.01\times$ speedup and $85.28\%$ energy savings, significantly boosting computational efficiency in digital SRAM-PIM systems.

\end{abstract}


\IEEEpeerreviewmaketitle

\section{Introduction}\label{sec:Introduction}
\begin{figure}[t]
  \centering
  \includegraphics[width=\linewidth]{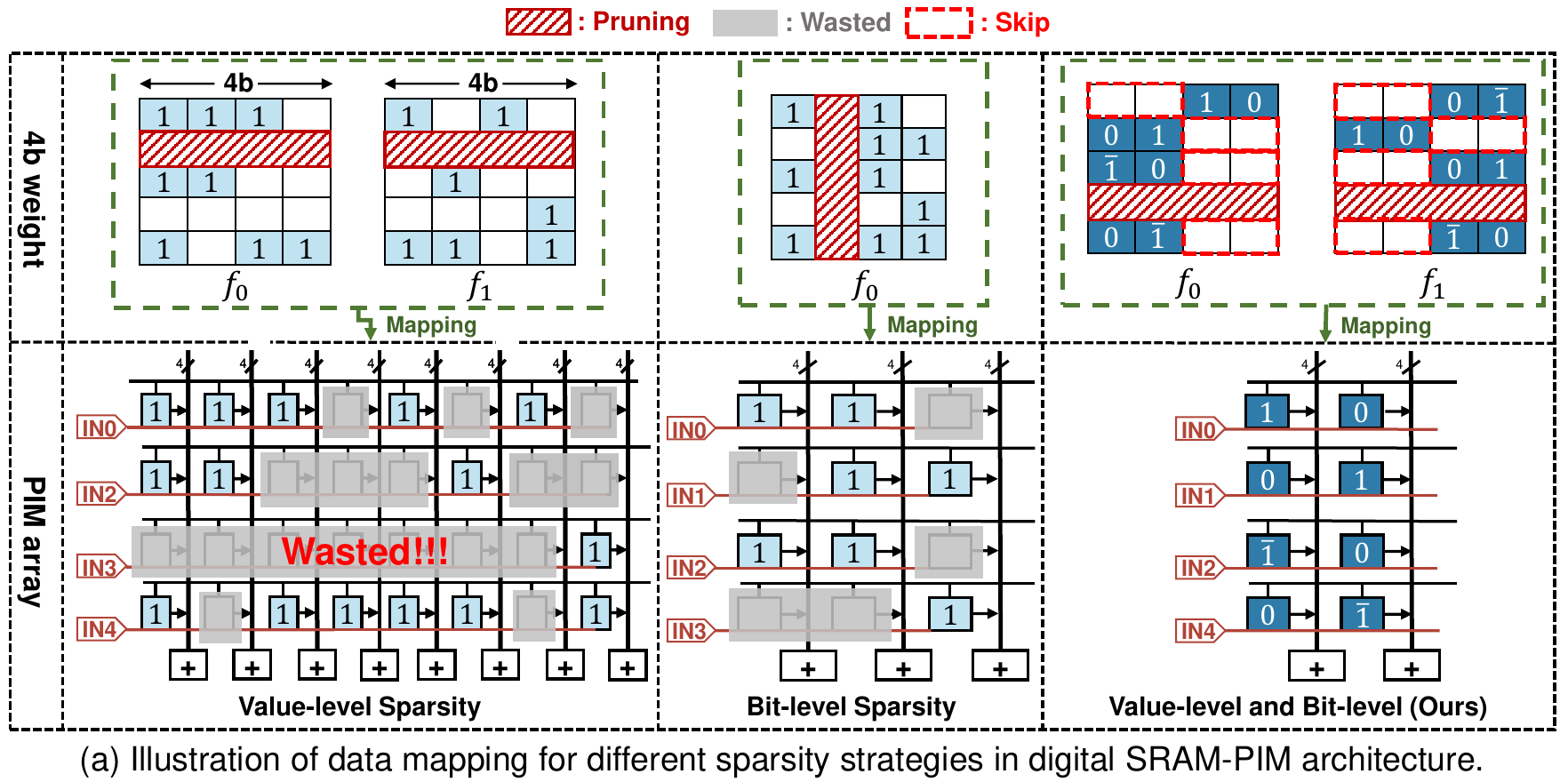}
  \vspace{-18pt}
\label{fig1:Comparison-a}
\end{figure} 
\begin{figure}[t]
  \centering
  \includegraphics[width=\linewidth]{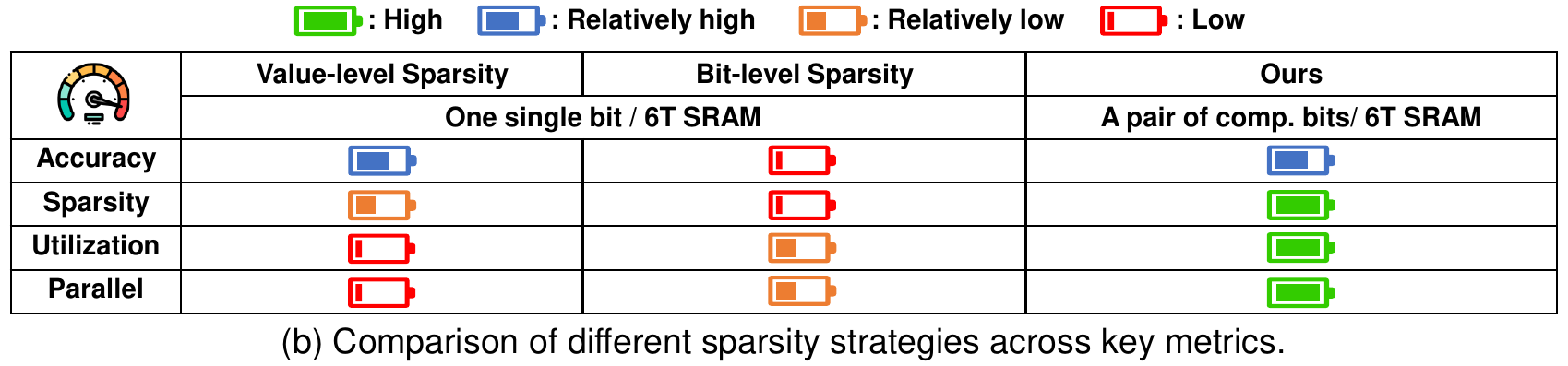}
  \vspace{-6pt}
  \caption{Comparative analysis of (a) data mapping strategies and (b) performance metrics (accuracy, sparsity, utilization, and parallel) for value-level sparsity, bit-level sparsity, and the proposed sparsity approach in digital SRAM-PIM architectures.}
\label{fig1:Comparison}
\end{figure}
\IEEEPARstart{D}{eep} Neural Networks (DNNs) have made remarkable advancements in a multitude of artificial intelligence (AI) applications, including image recognition, speech recognition, and object detection~\cite{Li_2022_CVPR,cheng2024yolo}.
The rapid growth of DNNs has led to a significant increase in model parameters and computational complexity.
These large-scale and compute-intensive models require substantial memory and computational resources, posing challenges to hardware performance and efficiency.
To mitigate these challenges, a range of compression techniques~\cite{dass2023vitality,li2023vit} have been proposed. 
Among these, sparsity exploitation~\cite{wei2023sparsifiner,yu2023boost} has emerged as a particularly effective approach, as it leverages the inherent redundancies in DNN models to optimize both memory access and computational efficiency.
For instance, analysis of neural networks (NNs) reveals a significant incidence of zero values, known as \texttt{value-level sparsity}, which results in numerous unnecessary computations. 
Skipping the computation of these zero values will significantly improve computational efficiency. 
In addition, a prevalent occurrence of zero bits within non-zero values, termed \texttt{bit-level sparsity}, offers further avenues for optimization. 
For example, the multiplication of an $INT8$ weight ($W$) and an input feature ($I$) can be decomposed into $64$ individual $1b \times 1b$ operations, as shown in Eq.~(\ref{eq1-mul}):
\begin{equation}
\label{eq1-mul}
I*W=\sum_{i=0}^7{\sum_{j=0}^7{I_{i} \times W_{j}}}.
\end{equation}
During the computation, only those pairings of non-zero $I_i$ and $W_j$ contribute meaningfully to the final results.
However, the proportion of non-zero bits in both weights and input features is relatively low, indicating that a significant majority of these computations are ineffectual.
Eliminating these redundant \texttt{zeros} (including zero values and zero bits) in storage and computation is paramount for accelerating DNNs.

In hardware, numerous DNN accelerators have also been proposed to exploit sparsity and tackle the computational demands of DNNs. 
For instance, Eyeriss V2~\cite{chen2019eyeriss} extends the flexible dataflow by introducing mechanisms to efficiently handle sparsity in both activations and weights, while SIGMA~\cite{9065523} introduces a Flexible Dot Product Engine (Flex-DPE), which can adapt to varying sparsity levels and dynamically map non-zero elements to its processing units.
However, these accelerators are predominantly based on the Von Neumann architecture, resulting in substantial energy consumption due to the frequent data movement between separate memory and computing units. 
Processing-in-memory (PIM) emerges as an innovative computational paradigm, offering a potential solution by executing multiply-accumulate (MAC) operations in memory.
Various memory technologies, such as SRAM \cite{fu2023p,duan2023ddc,qi2025cimflow}, RRAM \cite{lu2023rram,DBLP:journals/tcasI/MengWZL24}, and MRAM \cite{wang2020tcim,chen2022accelerating}, have been explored as potential candidates for PIM.
Among these, SRAM has gained widespread adoption in both academia and industry due to its high write speed, low energy consumption, and compatibility with established process technologies.

Unlike traditional digital accelerators, PIM-based accelerators are subject to strict data routing constraints due to their rigid crossbar structure.
This constraint introduces a \textbf{computational dependency issue} wherein computation can only be skipped if an entire block of data is zero.
Consequently, it impedes the efficient utilization of randomly distributed \texttt{zeros}.
This challenge is particularly pronounced in the highly parallel architecture of digital SRAM-PIM, which mandates the simultaneous activation of the entire array for computation.
Current SRAM-PIM research mainly focuses on dense NN acceleration or structured value-level sparsity support to enhance computation efficiency. 
Even though extensive efforts have been made to explore these sparsity techniques, they still fall short of harnessing the full potential of sparsity, as shown in Fig.~\ref{fig1:Comparison}.

Structured value-level sparsity that is friendly to digital SRAM-PIM architectures suffers from a relatively low compression ratio.
Schemes that target structured bit-level sparsity to capture finer-grained redundancy typically incur prohibitive accuracy degradation.
Additionally, the retention of \texttt{zeros} mapping within the PIM array results in the \textbf{low array utilization issue}.
Despite various studies proposing refined mapping strategies to enhance array utilization, the pervasive \texttt{zeros} continues to hinder optimal utilization.
Notably, the cross-coupled structure of 6T SRAM, along with digital in-memory customization features, offers a unique pathway for effectively utilizing unstructured bit-level sparsity. 
Moreover, unstructured bit-level sparsity combined with structured value-level sparsity can further eliminate redundant computation, which can maximize efficiency and performance.

To better quantify the efficiency of the PIM array, we define the \textit{actual utilization}, denoted by $\mathcal{U}_{act}$, as follows:
\begin{equation}
\label{eq1}
\mathcal{U}_{act}=\frac{EffectiveCompSRAMCells}{TotalCompSRAMCells} \times 100\%.
\end{equation}
This metric quantifies the proportion of SRAM cells actively engaged in effective computation relative to the total SRAM cells utilized per computation.
It serves as a critical measure for evaluating the efficiency of SRAM-PIM architectures in managing sparsity.

This paper is the journal extension version of our The Chip to System Conference (DAC) 2024 work \cite{duan2024towards}, providing comprehensive insights in response to the increasing complexity of modern NN models and computational workloads.
The most important extension is to fully exploit all available sparsity in NN models and modify the architecture to accommodate the diverse computational requirements.
The main contributions of this work are as follows:
\begin{itemize}
\item We propose the first algorithm and architecture co-design framework tailored for digital SRAM-PIM that effectively exploits value and bit sparsity in weights, as well as bit sparsity in input features.
This framework effectively addresses critical limitations and longstanding inefficiencies in sparsity utilization within digital SRAM-PIM architectures.
\item We introduce a hybrid-grained pruning algorithm that seamlessly combines coarse-grained structured value-level sparsity with randomly distributed bit-level sparsity in weights. 
This approach balances high compression rates and computational efficiency while preserving model accuracy, ensuring a practical and robust solution for accelerating sparse neural networks.
\item 
We design a customized DB-PIM architecture that incorporates multiple sparsity-aware architectural components to maximize the benefits of sparsity by eliminating the vast majority of superfluous computations.
\end{itemize}

The rest of this paper is organized as follows.
Sec.~\ref{sec: Motivation} provides background and motivations.
Sec.~\ref{sec: Overview} demonstrates an overview of the co-design framework. 
Sec.~\ref{sec: algorithm} and Sec.~\ref{sec: architecture} detail the proposed algorithm and architecture, respectively.
Experimental results are illustrated in Sec.~\ref{sec: Experiments}.
Discussion and conclusion are given in Sec.~\ref{sec:discussion} and Sec.~\ref{sec: Conclusion}.

\section{Background and Motivations}
\label{sec: Motivation}
\subsection{Richness of Sparsity in NN models}\label{sec:2.1}

\begin{figure}[t]
\centering
\includegraphics[width =1.0\linewidth]{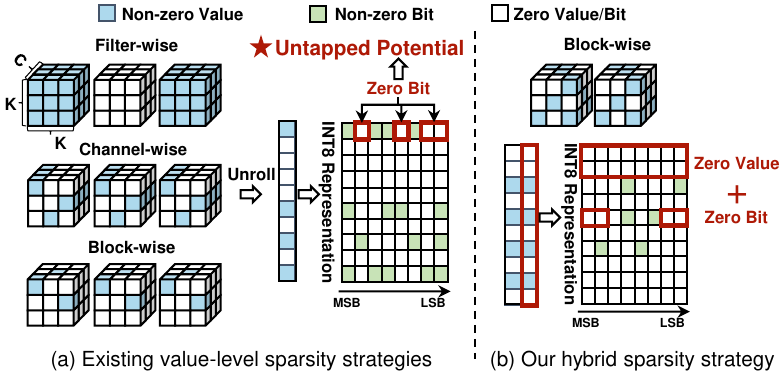}
\caption{Comparison of value-level sparsity and our proposed hybrid sparsity approach. (a) Illustration of existing value-level sparsity strategies and their limitations. (b) Emphasis on the advantages of our proposed hybrid sparsity method.}
\label{fig2:integration}
\end{figure}

The redundancy exhibited by NN models presents an opportunity to enhance efficiency by eliminating unnecessary storage and computation. 
This insight has catalyzed the development of numerous model compression techniques designed to exploit such redundancy for resource savings. 
Among these techniques, pruning has gained prominence because it systematically removes non-essential parameters while preserving model accuracy.
Since 2015, various pruning strategies have been explored, as shown in Fig.~\ref{fig2:integration}\hm{(a)}. 
Techniques such as filter-wise~\cite{wen2016learning} and channel-wise~\cite{liu2017learning} pruning have gained popularity due to their compatibility with various hardware architectures.
However, maintaining model accuracy with these methods typically constrains the compression ratio, resulting in limited acceleration effects.
This has led researchers to investigate more refined pruning techniques that offer higher compression rates without significantly impacting model accuracy.
Fine-grained pruning approaches, including block-wise~\cite{wen2016learning} and element-wise~\cite{sun2017meprop} pruning, have emerged as a promising solution. 
These methods provide a more nuanced control over which computations are omitted, allowing for a more targeted elimination of redundancies.
However, due to the presence of zero bits within non-zero values, the potential for computational efficiency gains has not yet been fully realized.
Therefore, we aim to combine value-level and bit-level sparsity to fully unlock this computational potential, as illustrated in Fig.~\ref{fig2:integration}\hm{(b)}.

\begin{figure}[t]
    \centering
    \subfigure[The proportion of zero bits in weights across different models. `Ori.' represents the original NN models, `Val.' indicates models with $60\%$ value-level sparsity, and `Our' refers to models with hybrid-grained sparsity ($60\%$ value-level sparsity and bit-level sparsity).]{
        \centering
        \includegraphics[width=0.48\textwidth]{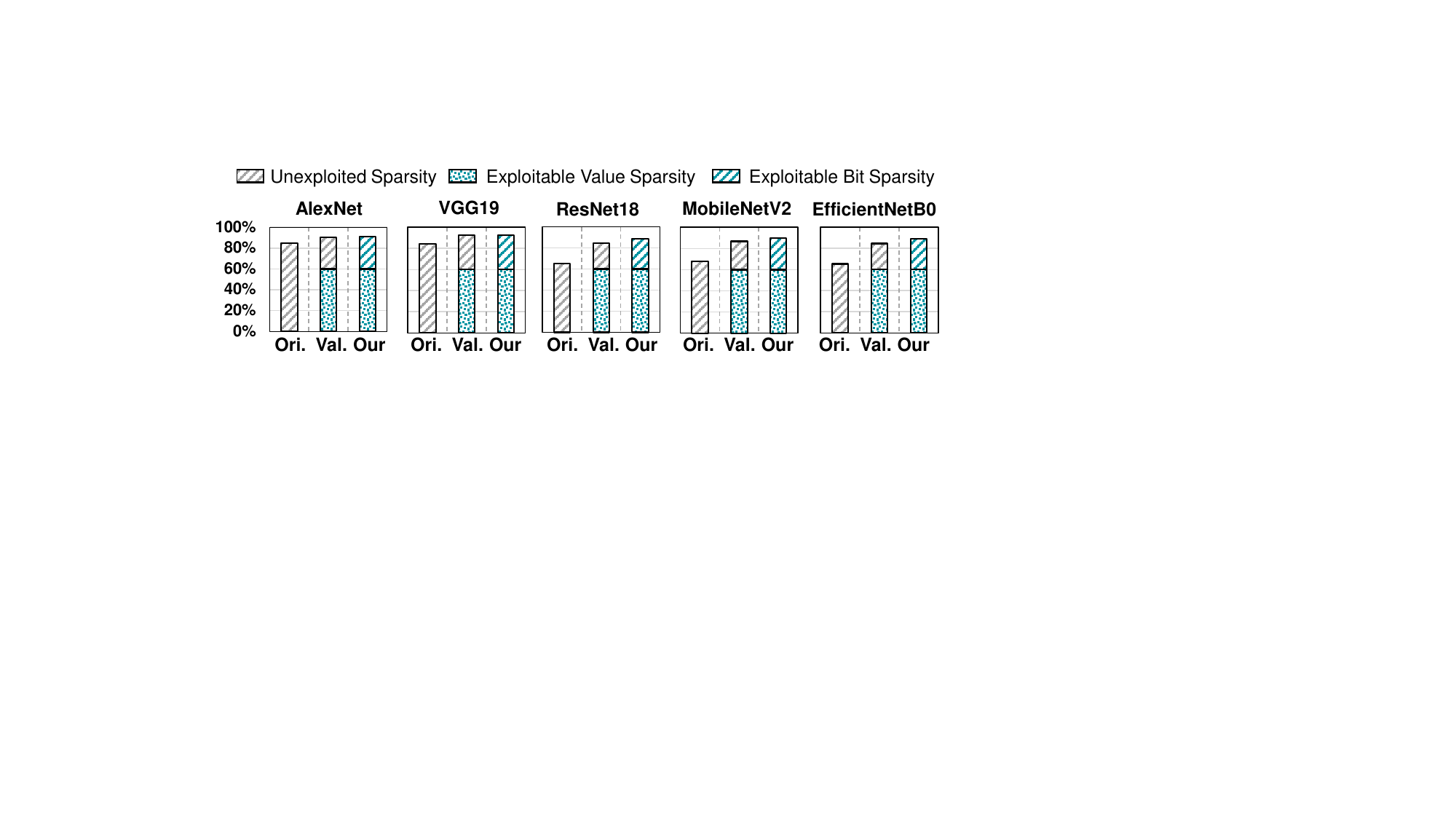}
        \label{fig:weight sparsity}
    }
    \subfigure[Proportion of all-zero columns (all zero in same bit position) in groups of N consecutive inputs (N = 1, 8, or 16). ]{
        \centering
        \includegraphics[width=0.48\textwidth]{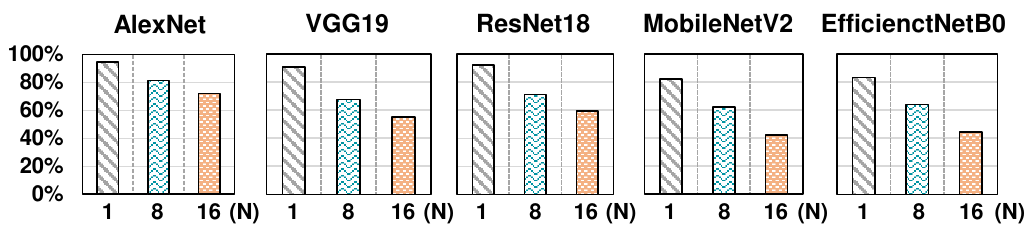}
        \label{fig: input sparsity}
    }
    \caption{Analysis of sparsity exploitation in weights and input features across various NN models.}
    \label{fig:sparsity}
\end{figure}

Building on this, we provide a data-driven perspective to analyze the sparsity exploitation across various NN models. 
As illustrated in Fig.~\ref{fig:weight sparsity}, after applying a block-wise pruning algorithm with $60\%$ value-level sparsity, the actual percentage of zero bits is higher than $80\%$.
This means that the proportion of zero bits within non-zero values is still substantial. 
Even for highly compressed NN models, these zero bits still offer opportunities for further efficiency improvements, which existing work often fails to exploit.
Therefore, we propose a hybrid pruning method that combines bit-level and value-level sparsity of weights to further improve the exploitable ratio of zero bits.
Bypassing these redundant calculations introduced by hybrid sparsity can significantly enhance computational efficiency and open new avenues for model optimization.

As illustrated in Fig.~\ref{fig: input sparsity}, input features also exhibit significant bit-level sparsity. 
However, bypassing all zero bits of input features in PIM presents practical challenges due to inherent architectural constraints.
Our analysis reveals that when input features are grouped into sets of 8 or 16, the probability that identical bit positions are zero across the group is relatively high (up to $80\%$ in groups of $8$, and around $70\%$ in groups of $16$). 
This observation suggests that exploiting this group-level sparsity can lead to substantial computational efficiency gains. 
\vspace{-8pt}
\subsection{Sparsity Exploitation in SRAM-PIM}\label{Sparsity Exploitation in SRAM-PIM}
In contrast to traditional architecture, PIM architecture integrates computation logic directly within the memory array to effectively address the bottleneck of frequent data movement between processing units and memory. 
This in-situ computation leverages the inherent parallelism and internal data bandwidth of memory to execute data-intensive operations like matrix-vector multiplications (MVM) directly within the memory cells.
For example, to perform MVM operations, weight matrices are first pre-loaded into memory cells.
In the analog domain, input vectors are streamed via wordline drivers and activate multiple rows simultaneously, enabling analog charge-domain MVM through conductance modulation in the memory cell.
In contrast, digital-domain PIM approaches broadcast input features into SRAM arrays, performing parallel bitwise AND operations between input and stored weights via embedded logic gates integrated into the memory array. 
Then, the bitwise computation results are accumulated through adder trees and shift\&and units~\cite{yan20221}.
 
To further boost computation efficiency, recent studies have focused on exploiting the sparsity of NN models on PIM architectures.
Yue et al.~\cite{yue202014} pioneered the implementation of block-wise zero-skipping in analog SRAM-PIM architectures, establishing a foundational approach to sparsity optimization. 
This strategy reduces power consumption by shutting down partial ADCs, however, it simultaneously reduces the utilization of computing resources as those ADCs cannot contribute to calculations.
In addition, all-zero sparse blocks still need to be stored in SRAM-PIM, further reducing the utilization of the array. 
To address the above problems, Yue et al.~\cite{yue202115} proposed a set-associate block-wise zero skipping architecture.
However, due to the ADC limitation, only part of the cell array is activated simultaneously. SDP~\cite{tu2022sdp} proposed a novel digital SRAM-PIM with a double-broadcast hybrid-grained pruning method, as well as a Bit-serial Booth in-SRAM (BBS) multiplication dataflow to further enhance efficiency.
Liu et al.~\cite{liu202316} proposed a zero skipper based on the butterfly network for unstructured NN models.

Based on the previous analysis in Sec.~\ref{sec:2.1}, the presence of zero bits in non-zero values limits the full potential of computational efficiency gains. 
A majority of zero bits are still required to be stored and processed in SRAM-PIM, resulting in a substantial number of ineffective calculations and low utilization.
The exploitation of bit-level sparsity could potentially overcome the above limitations.
Thus, some researchers have explored bit-level sparsity in SRAM-PIM architecture. 
Tu et al.~\cite{tu202316} develop the Bandwidth-Balanced CIM (BB-CIM) architecture to mitigate computational imbalances arising from input bit-level sparsity.
Similarly, TT$@$CIM~\cite{guo2022tt} aims to improve energy efficiency by designing a bit-level-sparsity-optimized CIM macro with $1$'s/$2$'s complement mode mixed computation. 
Rios et al.~\cite{10024996} introduces a novel approach, employing a hybrid quantization scheme combined with bitwise multiplication skipping to significantly improve inference efficiency, achieving substantial performance improvements while reducing computational overhead.
However, a common limitation underlies these pioneering efforts.
In highly parallel digital SRAM-PIM, the prerequisite for skipping a computation is that all bits within an entire computational column must be simultaneously zero. 
This requirement leads to significant accuracy loss and fails to eliminate all zero bits, thereby impeding optimal array utilization.

To solve this problem, we propose the DB-PIM co-design framework.
It aims to comprehensively exploit block-wise value-level sparsity and unstructured bit-level sparsity of weight, as well as the input block-wise bit-level sparsity, thereby maximizing computational efficiency. 
\textbf{By utilizing CSD encoding, a hybrid-grained pruning algorithm, a novel sparsity pattern, and the DB-PIM architecture, we ensure that only effective non-zero bits are stored and computed. This approach significantly reduces invalid computations, leading to enhanced efficiency and higher utilization.}
\vspace{-4pt}

\section{Overview of DB-PIM Framework}\label{sec: Overview}
\begin{figure}[t]
\centering
\includegraphics[width =1.0\linewidth]{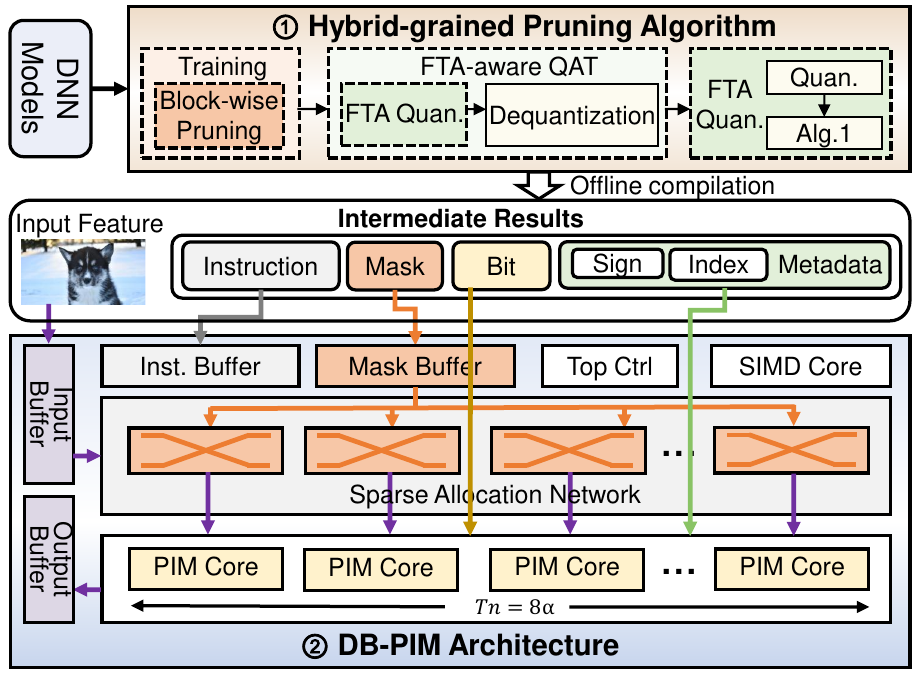}
\caption{An overview of the proposed DB-PIM, an algorithm and architecture co-design framework.}
\label{fig3:Overview}
\end{figure}

This section presents the overview of the proposed $\underline{\textbf{D}}$yadic $\underline{\textbf{B}}$lock $\underline{\textbf{PIM}}$ (DB-PIM), an algorithm and architecture co-design framework, as shown in Fig.~\ref{fig3:Overview}.
First, we propose a \textcircled{1} hybrid-grained pruning algorithm that combines structured value-level sparsity and unstructured bit-level sparsity.
This approach finds an optimal balance between these two types of sparsity, achieving high compression ratios while maintaining model accuracy.
We then design a dedicated \textcircled{2} DB-PIM architecture to support the above algorithm. 
This cohesive integration strategy effectively addresses the two fundamental issues previously identified.
Consequently, this improves hardware efficiency and increases array utilization.

In the training procedure, we first perform a coarse-grained pruning algorithm to obtain the pre-trained model.
Then, we apply a modified Quantization-Aware Training (QAT) to these models, known as Fixed-Threshold Approximation (FTA)-aware QAT.
This framework seamlessly integrates the FTA algorithm into the training loop, applying it in each epoch to obtain a fixed number of non-zero bits per filter. 
To support this process, we employ dynamic min-max quantization with exponential moving average (EMA) smoothing for range calibration. 
This approach eliminates reliance on precomputed global ranges and avoids introducing trainable parameters, thereby preserving computational efficiency.
To mitigate accuracy degradation, gradients are propagated using the straight-through estimator (STE), ensuring effective optimization under the imposed sparsity constraints.
This modification in QAT is crucial to obtain quantization parameters that reflect the impact of the FTA algorithm on model accuracy.
Finally, after fine-tuning, the model undergoes the final FTA quantization stage.
During this process, weights are reprojected to their nearest FTA-compliant quantized values, ensuring consistency with the algorithm constraints.

Following the training phase, an offline compilation transforms the FTA-quantized model into bits and metadata (e.g., signs and indices), and generates the corresponding masks and instructions tailored for our PIM architecture.
These intermediate results are then stored in off-chip memory.
At the hardware level, the DB-PIM architecture incorporates several sparsity-aware modules, explicitly optimized for our proposed hybrid-grained algorithm.
Key architectural innovations include sparse allocation networks and customized PIM macros designed to exploit both structured and unstructured sparsity.  
By strategically integrating these components, our approach minimizes unnecessary data processing, thereby significantly enhancing overall computational throughput and energy efficiency.

\section{Hybrid-grained Pruning Algorithm}\label{sec: algorithm}

\begin{figure*}[t]
\centering
\includegraphics[width =1.0\linewidth]{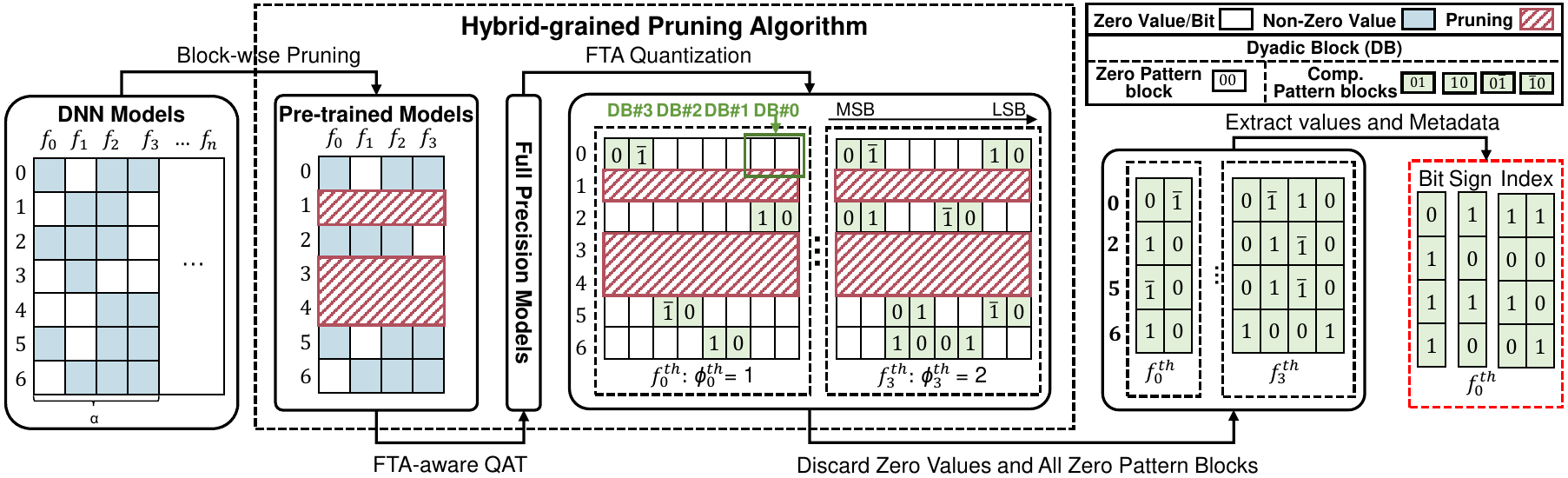}
\caption{Demonstration of hybrid-grained pruning algorithm.}
\label{fig4:Algorithm}
\end{figure*}

\subsection{CSD Encoding Scheme}\label{CSD_encoding}
In our work, we introduce a CSD encoding scheme~\cite{REITWIESNER1960231}, a ternary signed digit system, proposed by George W. Reitwiesner in 1960. 
This encoding scheme composed of $1$, $0$, and $-1$ (denoted as $\overline{1}$), to represent binary numbers.
CSD encoding exhibits three critical properties: 
\begin{enumerate}
    \item It has the least number of non-zero bits ($1$, $\overline{1}$).
    \item Adjacent bits cannot both be non-zero.
    \item CSD representation is unique.
\end{enumerate}

The CSD representation, on average, exhibits a 33\% reduction in non-zero bits compared to traditional binary equivalents~\cite{pinjare2013implementation}, thereby making it particularly well-suited for bit-level sparsity in hardware acceleration. 
This representation allows for efficient encoding of both positive and negative values without requiring an explicit sign bit or additional computational overhead.
Specifically, an 8-bit signed binary number, $W$, can be represented in an equivalent CSD representation as:
\begin{equation}
\label{eq2}
W = \sum_{i=0}^{7} w_{i} \times 2^{i}, w_{i} \in \left \{ -1,0,1 \right \}.
\end{equation}
For example, as shown in Tab.~\ref{Type}, the signed decimal number $-67$ can be encoded as $0\overline{1}00\_0\overline{1}01_{CSD}$, equivalent to:
\begin{equation}
    -2^6-2^2+2^0=-67.
\end{equation}
The conversion rule is that, from LSB to MSB, if consecutive 1s are found in binary representation (such as `$11$'), they are represented as $10\overline{1}$.

\begin{table}[]
\centering
\caption{Different types of number representation.}
\label{Type}
\begin{tabular}{|c|c|c|}
\hline
Decimal         & Two's Complement       &   CSD Representation                 \\ \hline
$67$            & $0100\_0011$            &   $0100\_010\overline{1}$             \\ \hline
$-67$         & $1011\_1101$            &   $0\overline{1}00\_0\overline{1}01$ \\ \hline
\end{tabular}
\end{table}

Incorporating the CSD encoding scheme into our framework stems from two crucial motivations. 
First, the utilization of the CSD encoding scheme substantially increases bit-level sparsity, thereby enhancing computational efficiency. 
Second, its inherent property of preventing consecutive non-zero bits is crucial for our DB-PIM architecture, as illustrated in Sec.~\ref{sec: architecture}.
This attribute fits seamlessly with the hardware constraints and significantly boosts the performance of the architecture, making it an optimal choice for our DB-PIM framework.

\subsection{Fine-grained Sparsity Pattern}\label{sec4.2}
Based on the CSD encoding scheme, we introduce a novel fine-grained sparsity pattern, called the dyadic block (DB). 
As highlighted in green in Fig.~\ref{fig4:Algorithm}, an 8-bit binary number can be partitioned into four DBs, `$DB\#3|DB\#2|DB\#1|DB\#0$', each consisting of a pair of bits. 
Each DB is assigned a distinctive index (e.g., `$\#0$') to denote its corresponding position, which is crucial for accurately locating non-zero bits. 
These DBs can be categorized into two types: Zero Pattern blocks ($00$) and Complementary (Comp.) Pattern blocks ($01$, $10$, $0\overline{1}$, and $\overline{1}0$).
To effectively quantify bit-level sparsity within our framework, we introduce a symbol $\phi$ to denote the number of non-zero bits.
It ranges from 0 to 4, corresponding to sparsity levels from $100\%$ down to $50\%$. 
For example, the 8-bit binary number $f_0^{th}(0) = 0\overline{1}00\_0000_{CSD}$ decomposes into four DBs: `$0\overline{1}|00|00|00$'. 
Among them, `$DB\#3$' is a Comp. Pattern block, while the others are Zero Pattern blocks.
Here, $\phi_0(0) = 1$ indicates only one non-zero bit in this value, corresponding to $87.5\%$ sparsity. 

The rationale for adopting this pattern stems from our discovery that, with CSD encoding, an 8-bit binary number only consists of Zero Pattern blocks and Comp. Pattern blocks.
Meanwhile, Comp. Pattern blocks are highly compatible with the cross-coupled structure in SRAM cells. 
Consequently, we can discard all Zero Pattern blocks and store Comp. Pattern blocks in a single 6T SRAM cell for parallel computation. 
This method not only retains the unstructured characteristic of bit-level sparsity but also leverages the cross-coupled structure to maximize the utilization of the SRAM-PIM array.
In this system, all SRAM cells involved in the computation are effectively utilized, addressing \textbf{the low array utilization issue}.
However, solely relying on this pattern is insufficient to resolve \textbf{the computational dependency issue}, which arises from the varying amount of non-zero bits in each weight. 
Indiscriminately eliminating all Zero Pattern blocks might result in irregularities in weight matrices.
Such irregularity conflicts with the rigid crossbar structure of the highly parallel digital SRAM-PIM.
To resolve this problem, we introduce the FTA algorithm, designed to maintain matrix regularity by capping the number of non-zero bits in each weight.

\begin{algorithm}[t]
\small
\KwIn{Quantized filters ${\mathcal{F}} \doteq \left[ f_0, \dots, f_{n-1} \right]$, where $f_{i} \in  D^{\mathcal{N}}$, $n$ is the number of filters, $\mathcal{N}$ is the number of weights in one filter, $D$ is determined by the quantization precision, e.g. $INT8$, Query Table $T(\phi^{th}) =\left \{ t \in D\mid \phi(toCSD(t)) = \phi_i^{th}\right \}$}
\KwOut{Approximation filters ${\mathcal{F}}^{th} \doteq \left[ f^{th}_0,\dots,f^{th}_{n-1} \right]$, filter thresholds $\Phi^{th} \doteq \left[ \phi^{th}_0,\dots,\phi^{th}_{n-1} \right]$.}
    \For{i \textbf{in} $[ 0, 1, 2, \dots,  n-1 ]$}{
        \For{j \textbf{in} $[0, 1, 2, \dots, {\mathcal{N}} - 1]$ }{
            $f^{csd}_i(j) \gets \text{toCSD}(f_{i}(j))$ \hfill{\textcolor{brown}{// CSD conversion}}\\
            $\phi_i(j) \gets \text{CountNonZeros}(f^{csd}_{i}(j))$ \hfill{\textcolor{brown}{// Count non-zero bits}} 
        }
        $m_i \gets Mode\left \{ 0\le j < \mathcal{N}\mid \phi_i(j)\right \}$ \hfill{\textcolor{brown}{// Compute the mode, excluding weights where the mask equals 0}}\\
        \uIf{$\forall j, \phi_i(j) ==0 $}{
            $\phi_i^{th} \gets 0$  \hfill{\textcolor{brown}{// All zero filter}} \\ 
           }
        \uElseIf{$m_i == 0$}{
           $\phi_i^{th} \gets 1$ \\
           }
        \uElseIf{$ 1\le m_i \le 2$}{
            $\phi_i^{th} \gets m_i$ \\
        }
        \uElseIf{$ m_i >  2$}{
            $\phi_i^{th} \gets 2$  \hfill{\textcolor{brown}{// Limit max threshold to 2}}
        }
        \For{j \textbf{in} $[0, 1, 2, \dots, {\mathcal{N}} - 1]$}{
           $f^{th}_i(j) \gets \underset {t \in T(\phi^{th}) } {\arg\min}\left | t - f_i(j) \right | $  \hfill{\textcolor{brown}{//  Closest num to $f_i(j)$ in $T$}}\\
    }
    }
\caption{Fixed Threshold ($\Phi_{th}$) Approximation}
\label{alg1}
\end{algorithm}

\subsection{Hybrid-grained Pruning Algorithm}\label{sec:hybrid-grained pruning algorithm}
Due to the rigid crossbar structure of the PIM macro, the corresponding computation can only be skipped if all weights in the same row are zeros.
Coarse-grained structured pruning can be easily adapted to fit this structure.
However, aggressively applying this pruning alone at high compression ratios (e.g., $90\%$) causes severe accuracy degradation.
To mitigate this, we propose a hybrid-grained pruning algorithm, as shown in Fig.~\ref{fig4:Algorithm}, that synergizes a coarse-grained pruning algorithm with a fine-grained FTA algorithm, enabling ultra-high compression ratios. 
For example, a $60\%$ value-level sparsity followed by $75\%$ FTA-driven bit-level sparsity achieves a compound compression ratio of $90\%$.
The hybrid-grained pruning algorithm consists of three stages: coarse-grained block-wise pruning, FTA-aware QAT, and FTA quantization.
\subsubsection{Coarse-grained Block-wise Pruning}
We employ coarse-grained block-wise pruning, and the detailed procedure is as follows.
First, the weight matrix of each layer is partitioned into multiple non-overlapped blocks, each of which contains the weights at the same position in multiple filters. 
The pruning granularity, denoted as $\alpha$, is determined by the number of columns in the SRAM macro and the filter threshold derived from the FTA algorithm.
In our design, $\alpha$ is set to $8$.
Next, the $L2$ norm is calculated for each block, and these blocks are subsequently sorted accordingly. 
The pruning threshold is established according to the specified sparsity level (e.g., a $50\%$ sparsity means pruning half of the blocks).
Blocks with $L2$ norms below this threshold are pruned, and corresponding masks are generated.
For example, as illustrated in Fig.~\ref{fig4:Algorithm}, the block size $\alpha$ is set to $4$.
The red sections represent pruned blocks, specifically showing that blocks $1$, $3$, and $4$ have been pruned. 
To mitigate potential accuracy loss, the pruned model undergoes fine-tuning, during which all pruned blocks are maintained at zero.

\subsubsection{FTA-aware QAT}\label{FTA-aware QAT}
We perform QAT in an FTA-aware manner to obtain quantization parameters of the above NN models.
In detail, we first quantize the coarse-grained pruned model into an 8-bit precision model.
Then, we apply the FTA algorithm to further increase the compression ratio through unstructured bit-level sparsity.
The core idea of the FTA algorithm involves setting a uniform threshold, denoted as $\phi^{th}$, for each filter.
This threshold ensures that the weights within each filter contain the same number of non-zero bits.
Although these non-zero bits are randomly distributed, this uniformity removes the same number of Zero Pattern blocks. 
As a result, weight matrices can be compressed into the regular structure, as illustrated in green in Fig.~\ref{fig4:Algorithm}. 
The processing of this algorithm, as shown in Alg.~\ref{alg1}, is detailed below:
\begin{enumerate}[label=\large\protect\textcircled{\small\arabic*}]
\item \textit{CSD Conversion}: We convert quantized $INT8$ filters into CSD representation. 
\item \textit{Threshold Computation}: Through analyzing the non-zero bits distribution of all weights in a filter, we ascertain a threshold $\phi^{th}$ for each filter.
Specifically, we compute the mode (the most frequently occurring value) of non-zero bits in the weight of each filter, excluding weights where the mask equals $0$.
In other words, the weights masked during coarse-grained block-wise pruning are excluded from the threshold calculation.
Analysis of the weight distribution in various NN models demonstrates that a $\phi^{th}$ value of 2 predominates across the filters. 
Therefore, to improve the sparsity while maintaining accuracy, we constrain $\phi^{th}$ to a range from $0$ to $2$.
This constraint not only ensures structured sparsity but also eliminates additional hardware storage overhead introduced by  CSD encoding. 
Under this constraint, an overall storage requirement does not exceed 8 bits per weight, effectively preventing an increase in memory usage.
For example, given $\phi_0 = \{2,0,1,0,0,1,3\}$, $mask_0 = \{1,0,1,1,0,1,1\}$, and $m_0 = 1$, $\phi_0^{th}$ is determined to be $1$.
\item \textit{Threshold Approximation}: The FTA algorithm sets the value of $f^{th}_i(j)$ as the closest value to $f_i(j)$ in set $T(\phi^{th})$. 
For example, given $f_0 = \{-63,0,64,0,0,-8,13\}$, with $\phi_0$ and $mask_0$ as defined in the previous paragraph, and a threshold $\phi_0^{th} = 1$. 
The transformed weight in $f_0$ is $\{-64,0,64,1,0,-8,16\}$.
Here, $f_0(1)$ and $f_0(4)$ remain unchanged at zero, as they were pruned by a coarse-grained block-wise pruning algorithm. 
In contrast, $f_0(3)$ has not been pruned but naturally has a value of zero. Hence, it can be modified by the threshold approximation step.
Furthermore, the number of non-zero bits for $f_0(0), f_0(3)$ and $f_0(6)$ are $\{2,0,3\}$, respectively.
Since these values do not match the threshold, the closest approximations to the original weights are determined from the available sets, ensuring that all approximations adhere to a threshold value of $2$.
\end{enumerate}

Finally, we perform de-quantization operations to ensure accurate gradient calculations and parameter updates.

\subsubsection{FTA Quantization} At the end of NN training, we perform FTA quantization to obtain $INT8$ filters with fixed non-zero bits in each filter. 

Our proposed FTA algorithm mitigates the bit-level irregularity by establishing a uniform $\phi^{th}$.
However, this potential remains underutilized in current SRAM-PIM architectures due to several limitations. 
First, the current PIM macro is incapable of executing parallel computations on the complementary states, denoted as $Q/\overline{Q}$, stored in the cross-coupled structure of 6T SRAM.
Second, the existing adder trees cannot directly accumulate outputs with randomly distributed non-zero bits.
Third, the current PIM macros do not support the functionality to skip hybrid-grained sparsity.
Hence, the DB-PIM architecture is explicitly designed to exploit the opportunities afforded by our hybrid-grained pruning algorithm, thereby enhancing computational efficiency.

\section{Overall Architecture Design of DB-PIM}\label{sec: architecture}

\begin{figure}[t]
\centering
\includegraphics[width = 0.9\linewidth]{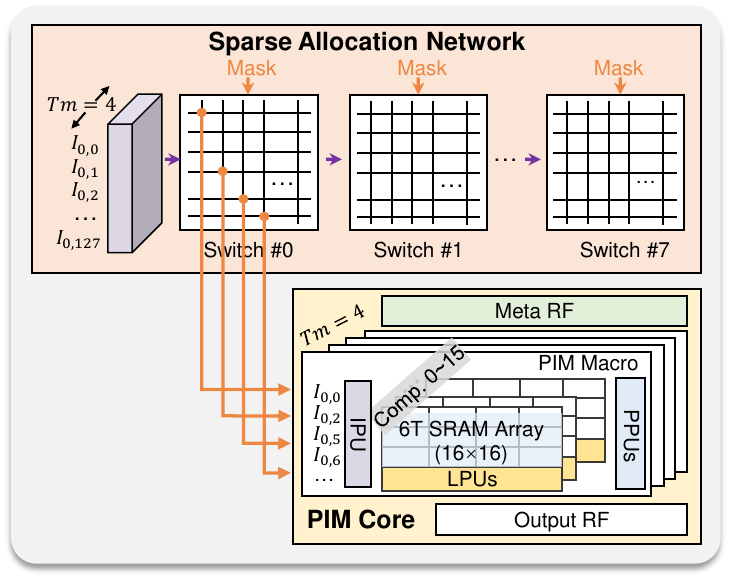}
\caption{Circuit design of sparse allocation network
and PIM core.}
\label{fig5:sparse-to-dense-b}
\end{figure}

\subsection{Top Level Architecture Design}\label{top level architecture design}

To support our hybrid-grained pruning algorithm and take advantage of multiple types of sparsity in digital SRAM-PIM, we propose a DB-PIM architecture, as illustrated in Fig.~\ref{fig3:Overview} \textcircled{2}.
It is composed of a top controller, a SIMD core, an instruction buffer, a mask buffer, an input buffer, an output buffer, a sparse allocation network, and eight homogeneous PIM cores.
Each PIM core, as shown in Fig.~\ref{fig5:sparse-to-dense-b}, consists of metadata register files (RFs) for storing signs and indices, four customized PIM macros for storing bits, and an output RF.
The PIM macro is an extension of the ADC-less SRAM PIM macro proposed in \cite{yan20221}.
To improve computational parallelism, all PIM macros in each PIM core store the same weights to compute different output features and processes in a pipelined manner.
The top controller first processes instructions retrieved from the instruction buffer (Inst. Buffer) and dispatches the corresponding control signals to the whole system. 
The input features retrieved from the input buffer are allocated to each PIM core through the sparse allocation network.
Each PIM macro only stores the Comp. Pattern blocks, skips all zero columns of input features, and executes bitwise \texttt{AND} operations.
The post-processing units (PPUs, including CSD-based adder trees, shift\&add units, and accumulators) are responsible for accumulating these results based on the metadata derived from the meta RFs and accumulation operations. 
Then, the results are written back into the output buffer.
Finally, the final MAC results are transferred into the SIMD core to perform other operations, such as depthwise conv, ReLU, pooling, element-wise multiply, quantization,.etc.

\subsection{Sparse Allocation Network}\label{sec:sparse allocation network}

\begin{figure}[t]
\centering
\includegraphics[width =\linewidth]{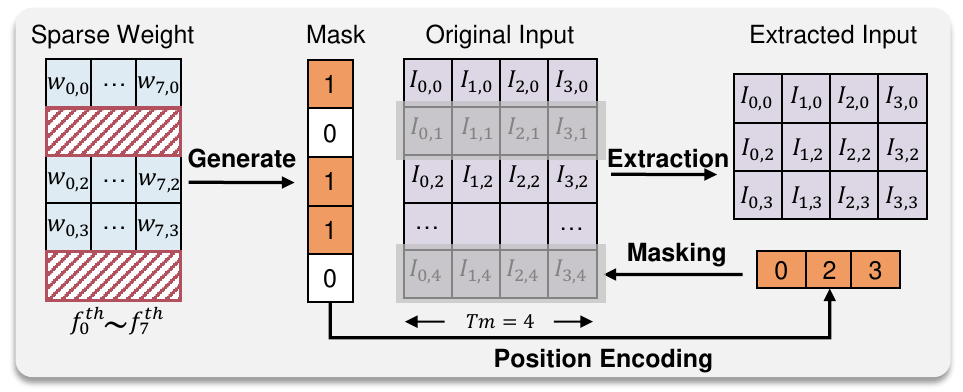}
\vspace{-16pt}
\caption{Illustration of the input extraction process.}
\label{fig5:sparse-to-dense-a}
\vspace{-6pt}
\end{figure}

Fig.~\ref{fig5:sparse-to-dense-b} illustrates the circuit design of a sparse allocation network, while Fig.~\ref{fig5:sparse-to-dense-a} details the input extraction process.
Initially, the weights are pruned using a block-wise pruning algorithm and generate corresponding masks.
These masks guide the selection of non-skipped input features, which are extracted from the original input based on mask positions.
The final products are obtained by multiplying the non-skipped weight and input features.
Specifically, the input features fetched from the input buffer are fed into a sparse allocation network. 
This network uses $8$ switches, using the leading-one detection module to extract inputs based on masks obtained from the mask buffer.  
This module scans the mask to identify the positions of the required input features, which are then used to control the MUX to select these inputs.
Then, the extracted input features from each switch are streamed to the corresponding PIM core for computation. 
To save input bandwidth, all switches share the same $128Tm$ input data. 
All macros within the same PIM core store identical weights and share one switch, which operates in pipeline mode to extract the input features sequentially.

\subsection{Customized SRAM-PIM Macro}

\begin{figure}[t]
\centering
\includegraphics[width = 1.0\linewidth]{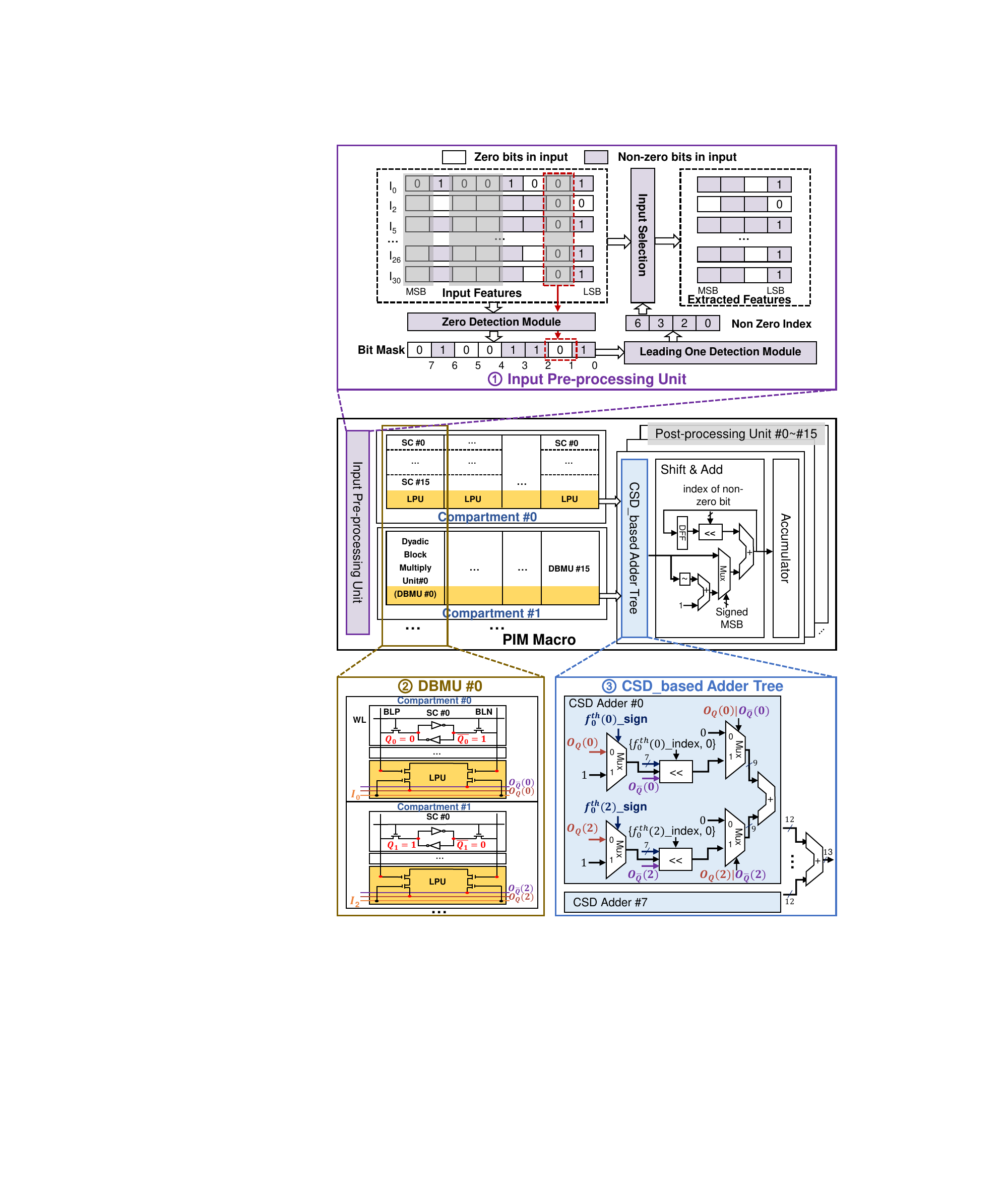}
\caption{Circuit design of customized SRAM-PIM macro.}
\label{fig6:post-processing}
\end{figure}

Fig.~\ref{fig6:post-processing} shows the circuit design of our customized SRAM-PIM macro, designed to exploit the bit-level sparsity both in input features and weights.
The macro consists of the IPU, an SRAM-PIM array with $16$ compartments, $16$ PPUs, and other peripheral circuits. 
The IPU is a bit sparsity-aware design adopted to dynamically detect and skip block-wise zero bits in input features.
SRAM-PIM array and PPUs are customized to skip the randomly distributed Zero Pattern blocks in the weight. 

In digital SRAM-PIM, the selected input features are sent to the PIM macro column by column and processed in a bit-serial manner.
Given that the number of non-zero bits varies across different input features, it is challenging to bypass all zero bits within the constraints of a rigid crossbar structure.  
However, as depicted in Fig.~\ref{fig: input sparsity}, block-wise zero bits continue to represent a substantial proportion of the overall data. 
Skipping the entire zero column can reduce the computation cycle required for processing the input feature.  
Thus, we propose IPU, as shown in Fig.~\ref{fig6:post-processing}\textcircled{1} to detect and skip block-wise zero bits dynamically.
Specifically, the input features, represented by $I_0$, $I_1$,...$I_{15}$, consist of both zero and non-zero bits.
Columns with all zero bits are shaded in gray, while non-zero bits are highlighted in purple.
The IPU first identifies columns consisting entirely of zero bits by the zero detection module and generates a corresponding bit mask. 
This mask is then employed by the leading one detection module to identify the non-zero indices within each input feature. 
The input selection module leverages these indices to selectively extract the relevant non-zero bits, effectively filtering out the all-zero columns and reducing the computation cycle that needs to be processed.

On the weight side, each compartment in the SRAM-PIM array comprises $16$ dyadic block multiply units (DBMUs), including sixteen 6T SRAM cells (SC \#0 $\sim$ SC \#15) and one local processing unit (LPU).
Each LPU within a DBMU consists of four transistors and acts as a fundamental dyadic block multiplier, performing two independent multiplications, $IN\times Q$ and $IN\times \overline{Q}$. 
In this setup, we can eliminate all Zero Pattern blocks and store a Comp. Pattern block in a cross-coupled structure ($Q$ and $\overline{Q}$) within the 6T SRAM cell of DBMU, executes two individual bitwise \texttt{AND} operations with identical input. 

For example, consider $f_0^{th}(0) = 0\overline{1}00\_0000$, $f_0^{th}(1) = 0000\_0000$ and $f_0^{th}(2) = 0000\_0010$, as shown in Fig.~\ref{fig4:Algorithm}. 
DB-PIM architecture eliminates block-wise zero values (such as $f_0^{th}(1)$) and randomly distributed Zero Pattern blocks and retains Comp. Pattern blocks with their corresponding indices and signs. 
We leverage the compatibility of Comp. Pattern blocks with 6T SRAM for parallel computation.
DB $\#3$ of $f_0^{th}(0)$ (e.g. 01) with $sign = 1$ and $index = 11$, and DB $\#0$ of $f_0^{th}(2)$ (e.g. $10$) with $sign = 0$ and $index = 00$ are stored into $Q_0 /\overline{Q_0}$ and $Q_1$ /$\overline{Q_1}$ in Fig.~\ref{fig6:post-processing}\textcircled{2} DBMU $\#0$. 
Subsequently, $I_1$ is bypassed, while $I_0$ and $I_2$ are directed to Compartment $\#0$ and $\#1$ for bitwise \texttt{AND} operations.
The equations are as follows:
\begin{align}
O_Q(0) = Q_0 \& I_0, O_{\overline{Q}}(0) = \overline{Q}_0 \& I_0, \\ O_Q(2) = Q_1 \& I_2, O_{\overline{Q}}(2) = \overline{Q}_1 \& I_2.
\end{align}
It should be noted that directly adding $\{O_Q(0)$,$O_{\overline{Q}}(0)\}$ and $\{O_Q(2)$,$O_{\overline{Q}}(2)\}$ would yield incorrect results.
For instance, if $I_0$ and $I_2$ are both $1$, the result by directly adding $\{O_Q(0)$,$O_{\overline{Q}}(0)\}$ and $\{O_Q(2)$,$O_{\overline{Q}}(2)\}$ would be $11$. However, the correct result is $f_0^{th}(0) + f_0^{th}(2) = 0\overline{1}00\_0000_{CSD} + 0000\_0010_{CSD} =1100\_0010_2$.
To address this, we have engineered a specialized CSD-based adder tree as shown in Fig.~\ref{fig6:post-processing}\textcircled{3}, specifically designed to handle randomly distributed non-zero bits.
The integration of our customized PIM macro and our algorithm allows us to efficiently exploit the structured block-wise bit-level sparsity of input features and the unstructured bit-level sparsity of weights in digital SRAM-PIM.

\subsection{Mapping Strategy}\label{sec:data mapping}

\begin{figure}[t]
\centering
\includegraphics[width = 1.0\linewidth]{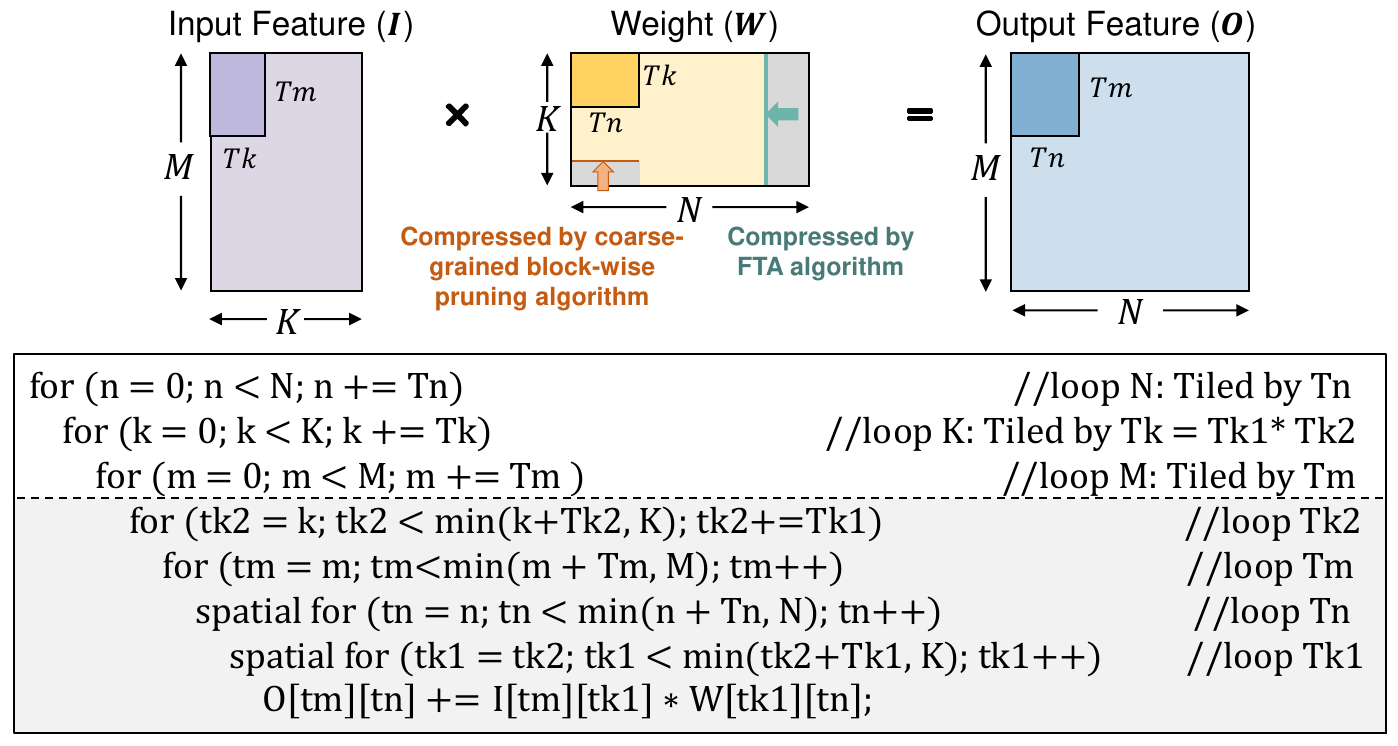}
\caption{Mapping strategy of NN models on the DB-PIM.}
\label{fig7:mapping}
\end{figure}

Due to the limited on-chip resources of SRAM-PIM, it is not feasible to map the weights of an entire layer simultaneously to the PIM array. 
Consequently, the layer must be partitioned into smaller tiles for processing. 
Specifically, we first convert three-dimensional convolutional operations into two-dimensional matrix multiplication operations through the $img2col$ function, as depicted below:
\begin{equation}
\label{eq7}
\textit{O}_{M\times N}=\textit{I}_{M\times K} \ast \textit{W}_{K \times N}.
\end{equation}
$M$, $N$, and $K$ represent the matrix dimensions of the unfolded input feature, weight, and output feature.
The weights are vertically compressed through the coarse-grained block-wise pruning algorithm and horizontally compressed through the FTA algorithm, as previously mentioned.
Consequently, only the compressed weights are stored in the PIM macros, rather than the entire weight matrices.

\begin{figure*}[t]
\centering
\includegraphics[width = 0.9\linewidth]{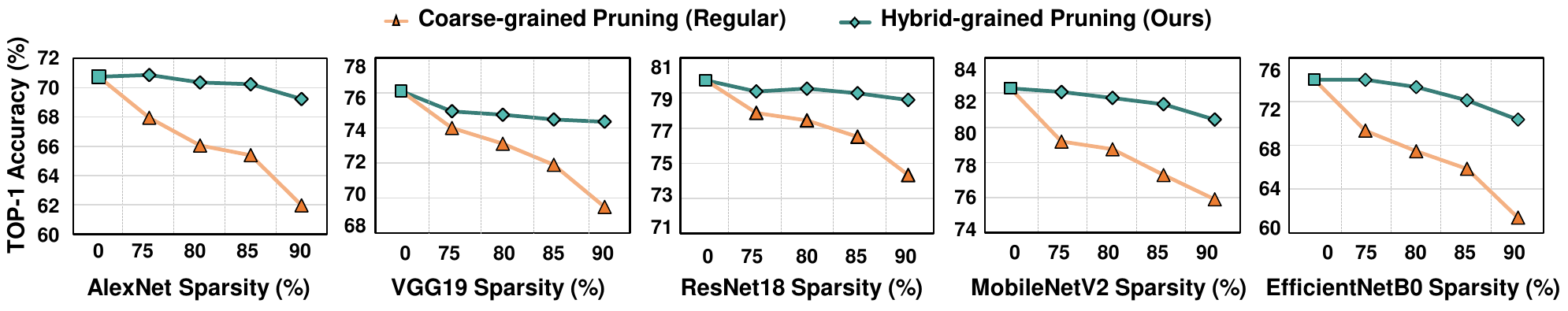}
\caption{Top-1 accuracy comparison in various NN models between our hybrid-grained pruning algorithm and coarse-grained pruning algorithm. Here, sparsity refers to bit-level sparsity, which can be exploited for computational efficiency and acceleration.}
\label{fig8:accuracy}
\end{figure*}
Fig.~\ref{fig7:mapping} illustrates the multi-level loop nest-based mapping strategy of DB-PIM for an entire layer of NN models.
The gray sections in pseudocode indicate the inner loop mapped onto the DB-PIM in parallel and remained stationary, while the rest form the outer loop scheduled by the controller.
Loop bounds are constrained by the configuration of hardware resources.
To minimize weight movement, DB-PIM employs a weight-stationary dataflow, ensuring that weights are preloaded into the SRAM array in the PIM macro and reused across multiple MAC operations without frequent re-fetching.
This approach exploits the high internal bandwidth of PIM and significantly reduces weight transfer overhead between on-chip and off-chip memory.
Specifically, the matrices are divided into several tiles by the parameters $Tm$, $Tn$, and $Tk$, where $Tk$ is further decomposed into $Tk1\times Tk2$.
The sizes of these tiles are optimized according to the capacity and parallel processing capability of the PIM macro to ensure efficient utilization of hardware resources.

Here, $Tm=4$ denotes the number of macros in each PIM core, with these macros storing identical weights to compute different output features, as discussed in Sec.~\ref{sec:sparse allocation network}. 
$Tn=8\alpha$ represents the number of output channels that can be calculated simultaneously within the $8$ PIM cores.
$\alpha$ is determined by the column count of each PIM macro and the threshold of the FTA algorithm.
$Tk1=16$ denotes the number of compartments within each macro, while $Tk2=16$ indicates the number of rows of SRAM cells in each compartment.
In the inner loop, $Tn$ and $Tk1$ of the loops can be spatially unrolled and computed simultaneously.
$Tk2$ of loops has to be computed sequentially, due to multiple rows of SRAM cells sharing a single LPU.
This structure allows multiple compartments to access the same row for parallel computation, followed by sequential traversal of different rows. During this process, partial sums are stored within accumulator registers, enabling output reuse and reducing redundant accesses to the output buffer.
The outer loop is executed in the sequence of $N-K-M$ to leverage the stationary nature of the PIM weights and release the output buffer storage.
This hierarchical and partitioned approach allows for efficient mapping and processing of neural network layers onto the limited resources of the SRAM-PIM architecture.

\section{Evaluation Results}\label{sec: Experiments}

\begin{figure*}[t]
\centering
\includegraphics[width =0.8\linewidth]{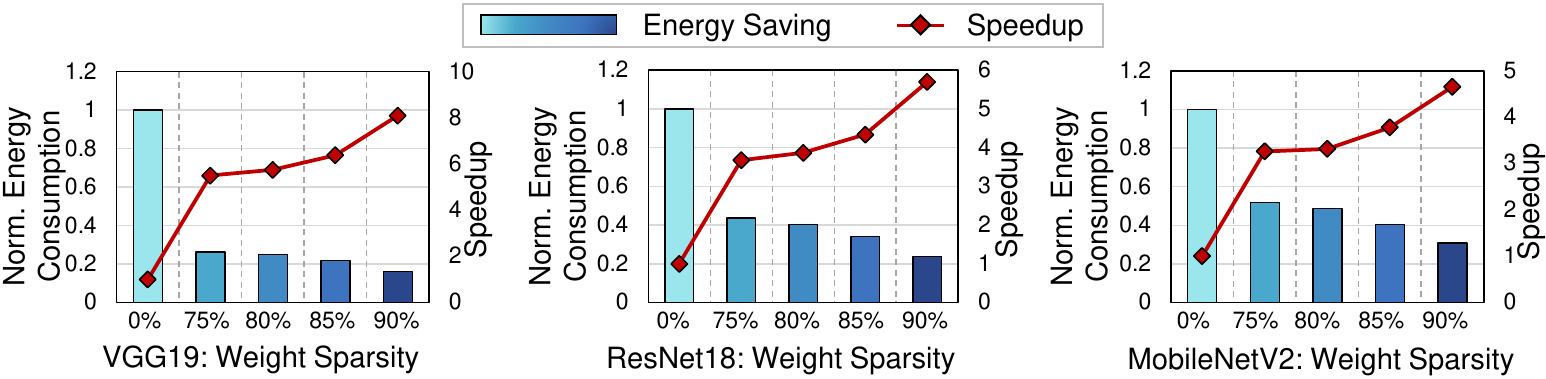}
\caption{Speedup and normalized energy consumptions over the dense PIM baseline on VGG19, ResNet18, and MobileNetV2 with hybrid-grained pruning algorithm at $75\% \sim 90\%$ sparsity.}
\label{fig10:breakdown analysis at sparsity-level}
\end{figure*}

\subsection{Experimental Setup}\label{setup}
\textbf{Training Protocol.} Models are trained for $500$ epochs using the AdamW optimizer, and the batch size is set to $128$.
Cosine annealing from $10^{-3}$ to $10^{-7}$, with $10$ epoch warmup ($10^{-5}$) and cooldown.
The accuracy results are strictly obtained under identical training budgets (such as epochs, batch size, and optimizer settings) for both the hybrid-grained pruning algorithm and baseline methods.

\textbf{Hardware Implementation.} DB-PIM is evaluated on $28$ nm technology, with a $128$ KB input buffer, a $256$KB output buffer, a $16$ KB instruction buffer, eight $2$ Kb mask RF, eight $1.5$ KB meta RF, and $16$ KB PIM capacity.
The parameters of DB-PIM are $Tm = 4$, $Tn = 8\alpha$, $Tk=Tk1\times Tk2=16\times16=256$.
The power consumption, latency, and area of the PIM macros are derived from post-layout analysis of customized design extension from \cite{yan20221}.
Meanwhile, the memory compiler is utilized to evaluate the area and power consumption of the involved memories.
The remaining digital modules are implemented with Verilog HDL and synthesized by Design Compiler for area evaluation, while PrimeTime PX is used to obtain power consumption. 
Aiming to evaluate the system functionalities and performance of DB-PIM, a compilation tool for dataflow mapping and a customized cycle-accurate C++ simulator are employed.

\textbf{Dense Digital PIM Baseline.} To evaluate the benefits and associated overhead of our proposed techniques, we established a baseline of dense digital PIM for comparison. 
This baseline, obtained by removing all sparsity support from the DB-PIM architecture, consists of buffers, the SIMD core, PIM cores, and the top controller. 
The PIM macro in the baseline is similar to the state-of-the-art digital PIM \cite{yan20221}, while all other hardware configurations are the same as in the DB-PIM.

\subsection{Accuracy Comparison of NN Models}\label{sec: Accuracy}
To demonstrate the effectiveness of our hybrid-grained pruning algorithm compared to a traditional coarse-grained pruning algorithm, we perform evaluations in various NN models.
Fig.~\ref{fig8:accuracy} illustrates the comparison results on three regular NN models such as AlexNet \cite{krizhevsky2012imagenet}, VGG19 \cite{simonyan2015very} and ResNet18 \cite{he2016deep}, along with two compact NN models such as MobileNetV2 \cite{sandler2018mobilenetv2} and EfficientNetB0 \cite{tan2019efficientnet}, on the CIFAR100 dataset. 
$INT8$ quantization is applied to both input features and weights in all layers.

For regular NN models, we apply coarse-grained block-wise pruning for \textit{standard convolution} (\texttt{std-conv}) layers and the FTA algorithm for the entire layers. 
Unlike regular NN models, compact NN models achieve efficiency by reducing parameter redundancy through structured decomposition.
For example, separable convolutions (\texttt{sp-conv}) decompose a standard convolution (\texttt{std-conv}) into depthwise (\texttt{dw-conv}) and pointwise (\texttt{pw-conv}) convolutions, significantly reducing computational cost while maintaining representational power \cite{deng2020model}.
\texttt{Dw-conv} layers generally lack adequate redundancy.
Thus, our hybrid-grained pruning techniques are mainly applied to \texttt{std-conv} layers and \texttt{pw-conv} layers, which perform most of the multiply-accumulate (MAC) operations in network calculations.

Our approach integrates structured value-level sparsity with unstructured bit-level sparsity. 
First, we apply block-wise value-level pruning with granularities of $20\%$, $40\%$, and $60\%$.
Subsequently, we utilize the FTA algorithm to introduce bit-level sparsity.
This algorithm dynamically thresholds filters between $0$ and $2$, producing bit-level sparsity levels ranging from $100\%$ ($\Phi^{th}$ === $0$, $\Phi^{th} \doteq \left[ \phi^{th}_0,\dots,\phi^{th}_{n-1} \right]$) down to $75\%$ ($\Phi^{th}$ === $2$). 
To standardize comparisons, we adopt $75\%$ as the baseline bit-level sparsity representation for all FTA threshold configurations in this paper, even when lower thresholds (e.g., between $0\sim2$) yield higher sparsity. 
For example, when $\Phi^{th} = \left[ 0,1,2,2,1,0\right]$, the sparsity of FTA algorithm reaches $87.5\%$, exceeding the minimum guaranteed $75\%$ sparsity.
This simple method ensures consistency across experiments while still reflecting the minimum guaranteed sparsity from the FTA.  
As shown in Fig.~\ref{fig8:accuracy}, $0$ sparsity refers to the baseline dense model without pruning.
While $75\%$ sparsity indicates bit-level sparsity introduced solely by the FTA algorithm.
An $80\%$ sparsity indicates the combined application of $20\%$ coarse-grained block-wise pruning along with the FTA algorithm.
By using the proposed hybrid-grained pruning algorithm, we can achieve over $90\%$ sparsity while keeping the accuracy loss within approximately $2\%$. 
In contrast, the traditional coarse-grained pruning algorithm results in an accuracy loss between $3\%$ and $5\%$ at $75\%$ sparsity and between $7\%$ and $12\%$ at $90\%$ sparsity.
Our hybrid-grained pruning method (depicted in green) consistently outperforms the coarse-grained pruning method (depicted in orange) across all NN models.
These results underscore the effectiveness of combining structured value-level sparsity with unstructured bit-level sparsity.

\begin{table*}[]
\caption{Detailed comparisons with related works.}
\vspace{-8pt}
\label{tab:comparison}
\resizebox{\textwidth}{!}{
\begin{threeparttable}
\begin{tabular}{|cc|c|c|c|c|c|ccccc|}
\hline
\multicolumn{2}{|c|}{}                       & \cite{yue202014} & \cite{yue202115} &\cite{kim2021z}  & \cite{tu2022sdp} & \cite{guo2022tt}  & \multicolumn{5}{c|}{This Work}  \\ \hline
\multicolumn{2}{|c|}{Technology (nm)}        & 65 & 65         & 65        & 28        & 28                & \multicolumn{5}{c|}{28}   \\ \hline
\multicolumn{2}{|c|}{Die Area (mm$^2$)}     & 5.66 & 8.32         &7.57             & 6.07                       & 8.97              & \multicolumn{5}{c|}{2.81}  \\ \hline
\multicolumn{2}{|c|}{Supply Voltage (V)}     & 0.9$\sim$ 1.05 & 0.62$\sim$1        & 1                 &     1 & 0.60$\sim$0.90      & \multicolumn{5}{c|}{0.72$\sim$0.80}     \\ \hline
\multicolumn{2}{|c|}{Frequency   (MHz)}      &50$\sim$100 & 25$\sim$100   & 200     & 500                        & 125$\sim$216      & \multicolumn{5}{c|}{333$\sim$500}      \\ \hline
\multicolumn{2}{|c|}{Power (mW)}         &  31.8$\sim$65.2  & 18.60$\sim$84.10   &5.29  & 1050               & 11.40$\sim$45.10    & \multicolumn{5}{c|}{28.92$\sim$281.26} \\ \hline
\multicolumn{2}{|c|}{SRAM Size (KB)}        & 164 & 294       &      4.75   & 384       & 114               & \multicolumn{5}{c|}{384}  \\ \hline
\multicolumn{2}{|c|}{PIM Size (KB)}      &  2  & 8                  & 38         &          128              & 128               & \multicolumn{5}{c|}{16}  \\ \hline
\multicolumn{2}{|c|}{Number of PIM Macro}  & 4 & 4          &8        & 512               & 16                & \multicolumn{5}{c|}{32}  \\ \hline
\multicolumn{2}{|c|}{Type}  & Analog & Analog        &Digital      & Digital           & Analog               & \multicolumn{5}{c|}{Digital }  \\ \hline
\multicolumn{2}{|c|}{\multirow{2}{*}{Dataset}}            & \multirow{2}{*}{\begin{tabular}[c]{@{}c@{}}MNIST/\\ CIFAR10\end{tabular}} & \multirow{2}{*}{\begin{tabular}[c]{@{}c@{}}CIFAR10/\\ ImageNet\end{tabular}} & \multirow{2}{*}{N/A$^\ast$} & \multirow{2}{*}{ImageNet} & \multirow{2}{*}{CIFAR10} & \multicolumn{5}{c|}{\multirow{2}{*}{CIFAR100}}          \\                                      \multicolumn{2}{|c|}{} &                    &                           &                 &        &                   & \multicolumn{5}{c|}{}      \\ \hline
\multicolumn{2}{|c|}{\textbf{Workload}}  &\textbf{ Conv-only}  & \textbf{Conv-only }          & \textbf{Conv-only}         & \textbf{\makecell{End-to-End \\ Evaluation}}               & \textbf{Conv-only   }             & \multicolumn{5}{c|}{\textbf{\makecell{End-to-End \\ Evaluation}}}  \\ \hline
\multicolumn{2}{|c|}{\multirow{2}{*}{\textbf{Actual Utilization} ($\mathcal{U}_{act}$)}}                                                  & ResNet18  & ResNet18   & VGG16        & ResNet-50        & \multicolumn{1}{c|}{\begin{tabular}[c]{@{}c@{}}ResNet20\end{tabular} } & \multicolumn{1}{c|}{AlexNet}      & \multicolumn{1}{c|}{VGG19}      & \multicolumn{1}{c|}{ResNet18}        & \multicolumn{1}{c|}{MobileNetV2}             &       EfficientNetB0                \\ \cline{3-12} 
\multicolumn{2}{|c|}{}  &\textbf{\textless32.04\%} &  \textbf{32.04\%} 
                                                & \textbf{16\% }   & \textbf{48.64\%} 
                                                     & \textbf{\textless50\%}   
                                                     & \multicolumn{1}{c|}{\textbf{85.04\%}} 
                                                     & \multicolumn{1}{c|}{\textbf{86.77\%}} 
                                                     & \multicolumn{1}{c|}{\textbf{86.29\%}}        
                                                     & \multicolumn{1}{c|}{\textbf{81.38\%}}   
                                                     & \textbf{78.44\%}      \\ \hline
\multicolumn{2}{|c|}{\textbf{Compression Ratio / Sparsity}}  &\textbf{$20\times$}& \textbf{$4.9\times$ }        
                                                               &\textbf{$90\%$}  & \textbf{$80\%$}  
                                                               & \textbf{N/A$^\ast$}  
                                                               & \multicolumn{5}{c|}{\textbf{$90\%$}}      \\ \hline
\multicolumn{2}{|c|}{\textbf{\makecell{Peak Throughput \\ (TOPS) (8b/8b)}}}  &\textbf{0.25 }& \textbf{0.10 }        
                                                               &\textbf{0.063}  & \textbf{26.21 }  
                                                               & \textbf{0.40  }  
                                                               & \multicolumn{5}{c|}{\textbf{2.48}}      \\ \hline
\multicolumn{2}{|c|}{\textbf{\begin{tabular}[c]{@{}c@{}}\makecell{Peak Throughput/Macro \\ (GOPS) (8b/8b)}\end{tabular}}}  & \textbf{62.5} & \textbf{24.69 } 
                                                                                                     & \textbf{7.95 } 
                                                                                                     & \textbf{51.19 }
                                                                                                     & \textbf{25.1 }       
                                                                                                     & \multicolumn{5}{c|}{\textbf{77.5}}   \\ \hline
\multicolumn{2}{|c|}{\textbf{\begin{tabular}[c]{@{}c@{}}\makecell{Peak Energy Efficiency per \\ Unit Area  (TOPS/W/mm$^2$) (8b/8b)}\end{tabular}}}       &     \textbf{0.79}  & \textbf{0.28}      
                                                                                                                      & \textbf{1.62} 
                                                                                                                      & \textbf{17.73} 
                                                                                                                      & \textbf{1.53 } 
                                                                                                                      & \multicolumn{5}{c|}{\textbf{2.37}}   \\ \hline
\end{tabular}
\begin{tablenotes}
    \item $\ast$ N/A indicates that the corresponding metric is not reported in the original paper.
\end{tablenotes}
\end{threeparttable}
}
\end{table*}

\subsection{Hardware Performance with Hybrid Sparsity}
Sec.~\ref{sec: Accuracy} provides a thorough assessment of our hybrid-grained pruning algorithm.
The experiments reveal that our algorithm markedly augments the compression ratio while maintaining high accuracy, surpassing the performance of regular coarse-grained pruning.
To further explore the benefits of our algorithms on hardware, we select VGG19, ResNet18, and MobileNetV2 to evaluate speedup and energy consumption across various sparsity levels.
Thus, we disable the dynamic skipping of block-wise zero columns in input features and focus solely on value-level and bit-level sparsity of weights. 
Meanwhile, we deliberately excluded operations such as ReLU, pooling, and depthwise convolution from the evaluation. 
This focused approach ensures a precise analysis of the algorithm's effectiveness in accelerating key computational layers. 
Fig.~\ref{fig10:breakdown analysis at sparsity-level} illustrates speedup and energy consumption over the digital PIM baseline at $75\% \sim 90\%$ weight sparsity. 
VGG19 attains a speedup of $5.50\times \sim 8.10\times$, while energy consumptions are reduced by $73.68\% \sim 83.90\%$ at weight sparsity between $75\% \sim 90\%$.
At $75\%$ sparsity, where only bit-level sparsity is employed, the speedup is roughly $4\times$ or higher.
This is because when the thresholds are set to $2$ for all NN models, a speedup close to $4\times$ is achievable.
However, in highly redundant network models, such as VGG19, filter thresholds vary between 0 and 2.
The actual bit-level sparsity introduced by the FTA algorithm significantly exceeds $75\%$, leading to a speedup that far exceeds $4\times$.
The hardware gains on ResNet18 and MobileNetV2 are relatively lower than those on VGG19. 
This is also attributed to the reduced redundancy in these models, leading to a higher proportion of weights with a threshold of $2$.
Furthermore, MobileNetV2 is worse than ResNet18.

\subsection{End-to-end Performance Breakdown}\label{end-to-end}

\begin{figure}[t]
    \centering
    \subfigure[Speedup achieved by different sparsity approaches across various NN models compared to the dense PIM baseline.]{
        \centering
        \includegraphics[width=0.48\textwidth]{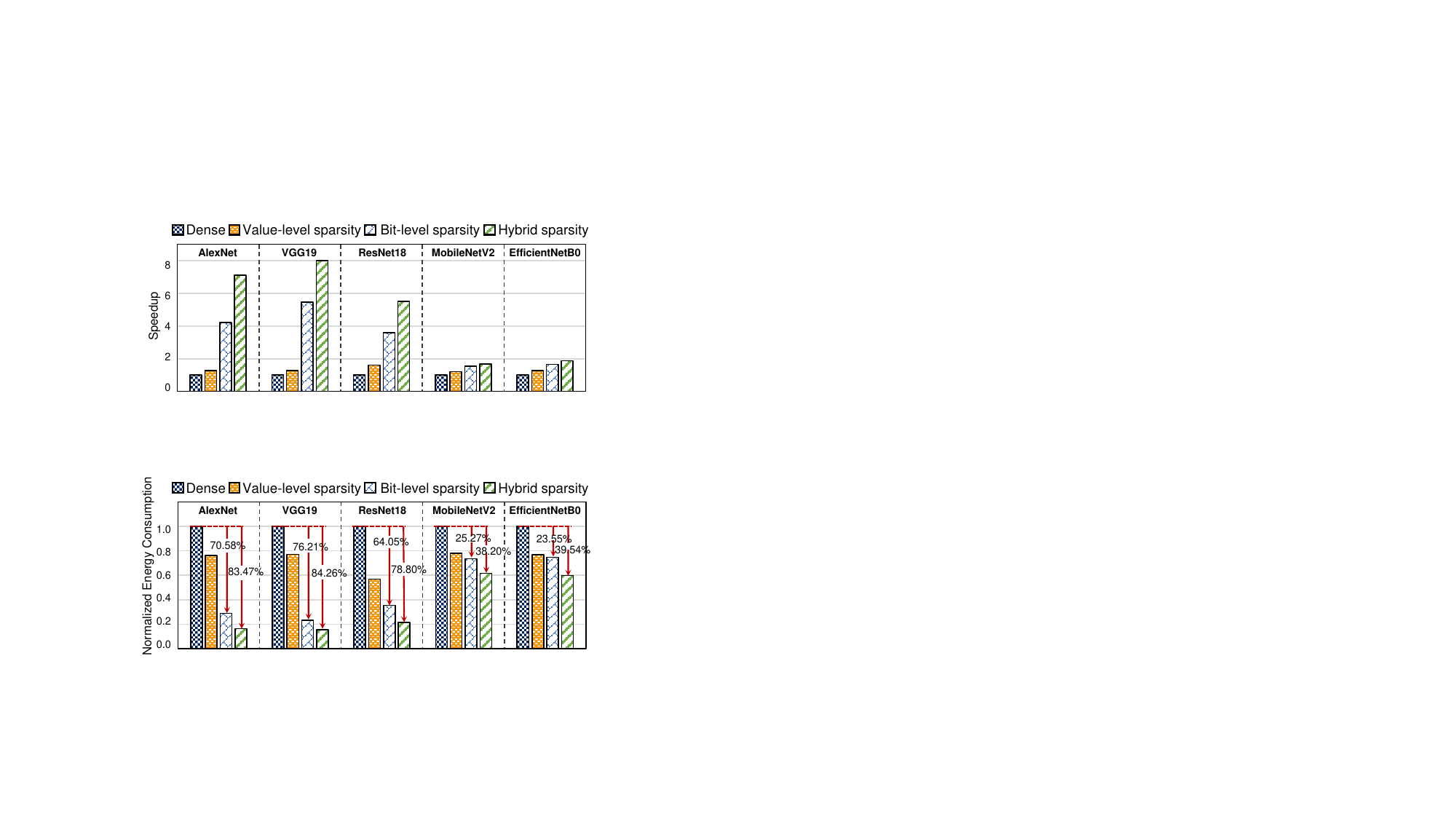}
    }
    \subfigure[Energy consumption normalized over the dense PIM baseline achieved by different sparsity approaches across various NN models.]{
        \centering
        \includegraphics[width=0.48\textwidth]{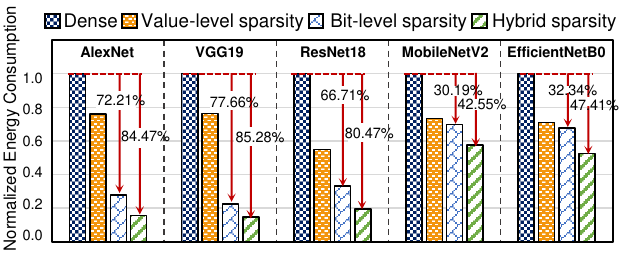}
    }
    \caption{Breakdown analysis for (a) speedup and (b) normalized energy consumptions achieved by different sparsity exploration approaches (bit-level sparsity of weights and input features, value-level sparsity of weights, and hybrid sparsity) across various NN models, compared to the dense PIM baseline.}
    \label{fig11:breakdown analysis at different strategy}
\end{figure}

\begin{figure}[t]
\centering
\includegraphics[width = \linewidth]{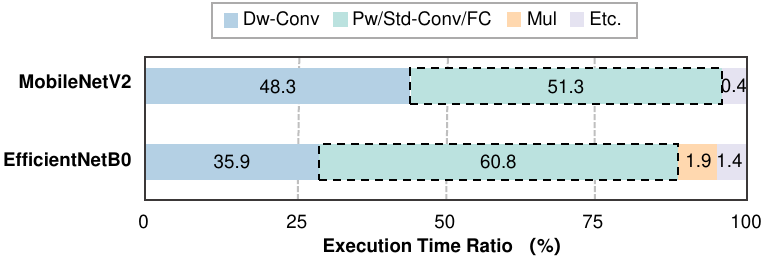}
\caption{The execution time breakdown of different operations in MobileNetV2 and EfficientNet.The proportions of depthwise convolutions (dw-Conv), pointwise/standard convolutions and fully connected layers (pw/std-Conv/FC), multiplications (Mul), and other operations (Etc., including pooling, ReLU, residual additions, and others) are shown.}
\label{fig12: Execution time}
\end{figure}

To further assess the hardware benefits of various sparsity exploration approaches, we present a detailed breakdown analysis of speedup and energy consumption relative to the digital PIM baseline. 
The analysis is performed on the end-to-end inference of complete neural network models rather than isolated layers, providing a comprehensive evaluation of the impact of the sparsity strategy.
It highlights the distinct contributions of different sparsity exploration techniques, including bit-level sparsity in weights and input features, value-level sparsity in weights, and hybrid sparsity, to overall system performance.
As illustrated in Fig.~\ref{fig11:breakdown analysis at different strategy}, employing bit-level or value-level sparsity individually yields significant improvements in speedup and energy efficiency. 
However, relying solely on one of these approaches does not achieve the optimal result.
By integrating techniques into a hybrid sparsity method, DB-PIM achieves optimal results by skipping zero values, zero bits within non-zero values, and dynamically bypassing all-zero columns in the input. 
This holistic approach maximizes the utilization of sparsity, unlocking substantial gains in performance and efficiency.

However, as illustrated in Fig.~\ref{fig11:breakdown analysis at different strategy}, MobileNetV2 and EfficientNet exhibit significantly lower speedups and energy consumption compared to other networks.
Thus, we present a detailed breakdown analysis of execution time in MobileNetV2 and EfficientNet.
As shown in Fig.~\ref{fig12: Execution time}, the execution time of \texttt{std/pw-conv} and FC layers in both compact models is only $51.3\%$ and $60.8\%$, respectively.
Dw-Conv ($48.3\%$ in MobileNetV2, $35.9\%$ in EfficientNet), multiplicative layers ($/$, $1.9\%$), and other non-linear computations (including pooling, ReLU, Residual addition, total $0.4\%$ and $1.4\%$ respectively) accounted for a significant portion of the total execution time.
However, the proposed DB-PIM framework primarily targets \texttt{std/pw-conv} and FC layers.
As a result, the high proportion of these additional operations significantly constrains the overall acceleration potential of these two models.

\subsection{Comparison with Prior Works}\label{sec:comparison with other works}
In previous sections, we conducted a thorough analysis of DB-PIM, covering both its algorithmic and hardware aspects. 
Here, we present a comprehensive comparison with existing state-of-the-art (SOTA) SRAM-PIM accelerators, which can be categorized into value-level sparsity \cite{yue202014,yue202115,tu2022sdp,kim2021z} and bit-level sparsity \cite{guo2022tt}. 
Tab.~\ref{tab:comparison} provides a summary of our work, including a total area, power consumption, and frequency.
The comparison mainly focuses on two critical performance metrics: utilization ($\mathcal{U}_{act}$) and peak throughput per macro.

\textbf{Utilization ($\mathcal{U}_{act}$):} 
As detailed in Sec.~\ref{sec:Introduction}, $\mathcal{U}_{act}$ adopted here specifically refers to the fine-grained bit-level utilization of SRAM-PIM, calculated by excluding all \texttt{zeros}.
Consequently, for prior studies, $\mathcal{U}_{act}$ here can be estimated through statistical analysis of non-zero bit ratios in NN models, serving as a theoretical upper bound of bit-level utilization.
Since mapping zeros in SRAM-PIM inevitably consumes resources, the actual utilization rates ($\mathcal{U}_{act}$) are often significantly below $50\%$.
Moreover, these theoretical values do not even account for utilization degradation incurred during deployment, such as dataflow mapping constraints and PIM macro allocation.

In contrast, our utilization values are derived from actual execution results on our architecture, providing a more accurate assessment of computational resource efficiency during inference. 
Leveraging our technique of storing two complementary bits in each SRAM cell, we ensure that every cell is effectively utilized during each computation.
As a result, $\mathcal{U}_{act}$ achieves an exceptionally high level, with values reported in \cite{duan2024towards} ranging from $94.41\%\sim98.42\%$ across various NN models. 
Furthermore, while architectural enhancements and the exploitation of bit-level weight sparsity significantly improve filter-level parallelism, we observe that in some convolutional layers, the number of output channels can be considerably smaller than the available computational parallelism. 
This mismatch limits the full utilization of PIM resources, subsequently reducing overall utilization.
As a result, our modified architecture achieves slightly lower actual utilization (approximately $80\%$) compared to previous implementations \cite{duan2024towards}.
Nonetheless, it remains highly competitive with state-of-the-art methods.

\textbf{Peak Throughput per Macro:}
Meanwhile, our approach offers superior theoretical peak throughput per macro compared to prior work. 
For example, conventional architectures employing $INT8$ quantization typically enable a $16$ column macro to process only two filters in parallel per cycle. 
In contrast, our method can process up to $16$ filters concurrently with a threshold of $1$ and $8$ filters with a threshold of $2$.
The theoretical peak throughput per macro is defined under fully dense workloads and is governed exclusively by architectural characteristics, irrespective of the dataset or model sparsity.
Under equivalent conditions, our approach theoretically achieves significantly higher peak throughput per macro than prior designs. 
Nevertheless, variations in macro sizes and operational frequencies across different implementations result in actual throughput values reported in Tab.~\ref{tab:comparison} falling short of the theoretical peak.

\section{Discussion}\label{sec:discussion}

\begin{table}[]
\centering
\caption{Detailed comparisons of on-chip execution time with our previous works$^{\ast}$.}
\label{excution}
\resizebox{\linewidth}{!}{
\begin{threeparttable}[b]
\begin{tabular}{|c|c|c|c|c|c|}
\hline
                            & AlexNet & VGG19 & ResNet & MobileNetV2 & EfficientNetB0 \\ \hline
DAC'24\cite{duan2024towards} (ms)     & 8.63    & 17.22 & 21.77  & 18.20       & 2.51           \\ \hline
Bit-level  (ms)                      & 2.88    & 4.37  & 4.03   & 2.34        & 0.40           \\ \hline
Hybrid-level (ms)                     & 1.69    & 2.96  & 2.60   & 1.64       & 0.30           \\ \hline
\end{tabular}
\begin{tablenotes}
    \item $^{\ast}$ The execution time includes only std/pw-conv and FC layer operations.
\end{tablenotes}
\end{threeparttable}
}
\end{table}
In this paper, we have made several architectural modifications compared with \cite{duan2024towards} to enhance performance and adapt to the diverse computational demands of modern NN models.
Firstly, to better support value-level weight sparsity resulting from block-wise pruning, we have introduced a sparse allocation network that efficiently skips computations involving zero values.
This network dynamically identifies and bypasses non-contributory input features, thereby reducing unnecessary computational overhead and optimizing processing efficiency.
Secondly, in response to the increasing complexity of modern neural network models and computational workloads, we have expanded the architecture to increase computational parallelism.
To further enhance functionality, we have upgraded the instruction set and refined a vector computational unit capable of supporting various non-linear operations such as pooling, quantization, ReLU, and residual additions (ResAdd).
These additions enable our architecture to efficiently handle end-to-end inference of complete NN models, as opposed to being limited to convolutional computations.  
As shown in Tab.~\ref{excution}, compared to our previous work, we can achieve up to $11.10\times$ speedup across various NN models.
\section{Conclusion}\label{sec: Conclusion}
This study presents Dyadic Block PIM (DB-PIM), an innovative algorithm-architecture co-design framework that effectively exploits multiple types of sparsity in digital SRAM-PIM systems.
By applying a block-wise pruning algorithm, DB-PIM efficiently bypasses computations associated with structured value-level sparsity.
Subsequently, our proposed FTA algorithm, combined with a novel sparsity pattern and the customized SRAM macro, are utilized to bypass unstructured bit-level sparsity within non-zero weights. 
Finally, the IPU is designed to dynamically skip the block-wise all-zero columns of input features.
Our results demonstrate that DB-PIM achieves up to $5.46\times$ speedup and $77.66\%$ energy savings by leveraging the bit-level sparsity of weights and input features. 
Moreover, when combined with structured value-level sparsity, DB-PIM attains even more remarkable results, with an $8.01\times$ speedup and an $85.28\%$ increase in energy efficiency.
These findings demonstrate that DB-PIM overcomes the inherent challenges of efficiently leveraging sparsity in digital SRAM-PIM.

{
\small
\bibliographystyle{IEEEtran}
\bibliography{ref.bib}
}

\vspace{-8mm}

\begin{IEEEbiography}[{\includegraphics[width=1in,height=1.25in,clip,keepaspectratio]{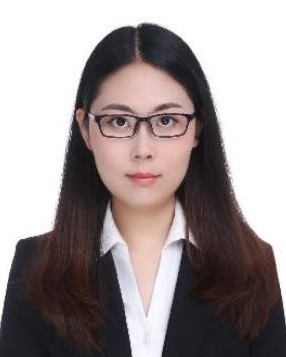}}]{Cenlin Duan}

received the B.S. degree in Electronic Science and Technology from University of Electronic Science and Technology of China, Chengdu, China, in 2015, and the M.S. degree in Software Engineering from Xidian University, Xi'an, China, in 2018. She is currently pursuing the Ph.D. degree at the School of Integrated Circuit Science and Engineering, Beihang University, Beijing, China. Her current research interests include processing-in-memory architectures and deep learning accelerators.

\end{IEEEbiography}

\vspace{-8mm}

\begin{IEEEbiography}[{\includegraphics[width=1in,height=1.25in,clip,keepaspectratio]{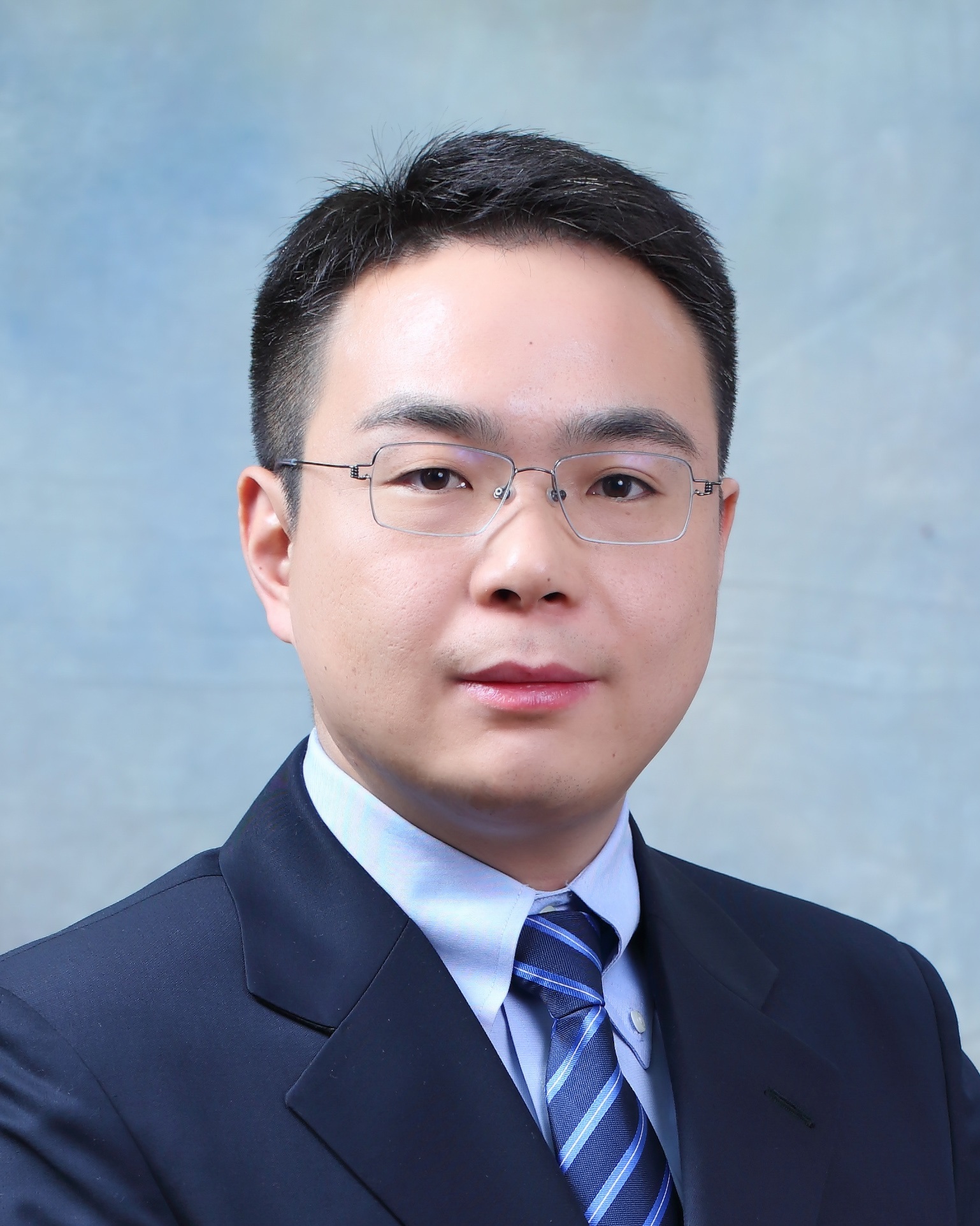}}]{Jianlei Yang}

(S'11-M'14-SM'20) received the B.S. degree in microelectronics from Xidian University, Xi'an, China, in 2009, and the Ph.D. degree in computer science and technology from Tsinghua University, Beijing, China, in 2014.

He is currently a Professor in Beihang University, Beijing, China, with the School of Computer Science and Engineering. From 2014 to 2016, he was a post-doctoral researcher with the Department of ECE, University of Pittsburgh, Pennsylvania, USA.
His current research interests include emerging computer architectures, hardware-software co-design and machine learning systems.

Dr. Yang was the recipient of the First/Second place on ACM TAU Power Grid Simulation Contest in 2011 and 2012. He was a recipient of IEEE ICCD Best Paper Award in 2013, ACM GLSVLSI Best Paper Nomination in 2015, IEEE ICESS Best Paper Award in 2017, ACM SIGKDD Best Student Paper Award in 2020.

\end{IEEEbiography}

\vspace{-8mm}

\begin{IEEEbiography}[{\includegraphics[width=1in,height=1.25in,clip,keepaspectratio]{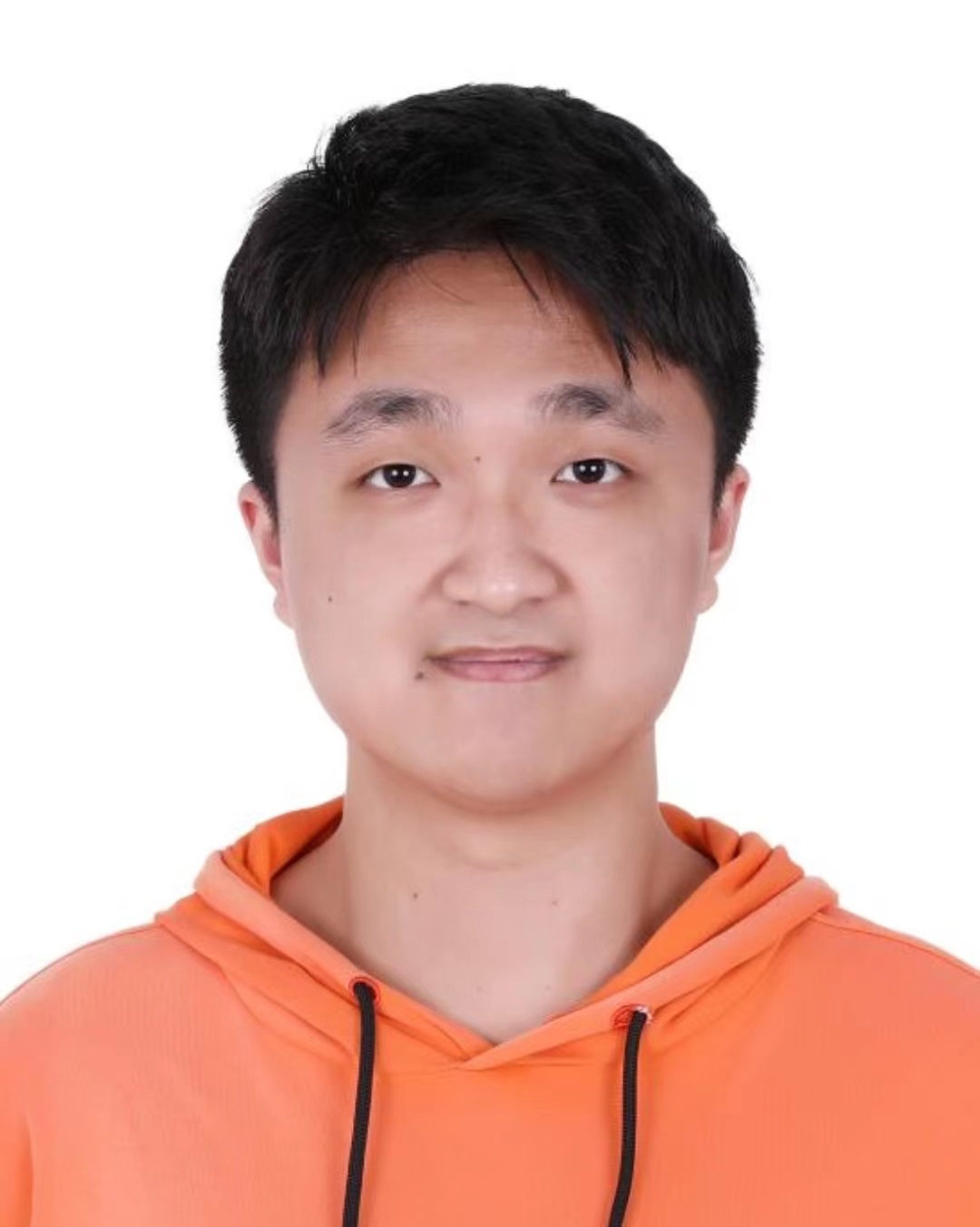}}]{Yikun Wang}

received the B.S. degree in computer science and technology from Beihang University, Beijing, China, in 2022. He is currently working toward the M.S. degree at the School of Computer Science and Engineering, Beihang University, China. His current research interests include computing-in-memory architectures and deep learning accelerators.

\end{IEEEbiography}

\vspace{-8mm}

\begin{IEEEbiography}[{\includegraphics[width=1in,height=1.25in,clip,keepaspectratio]{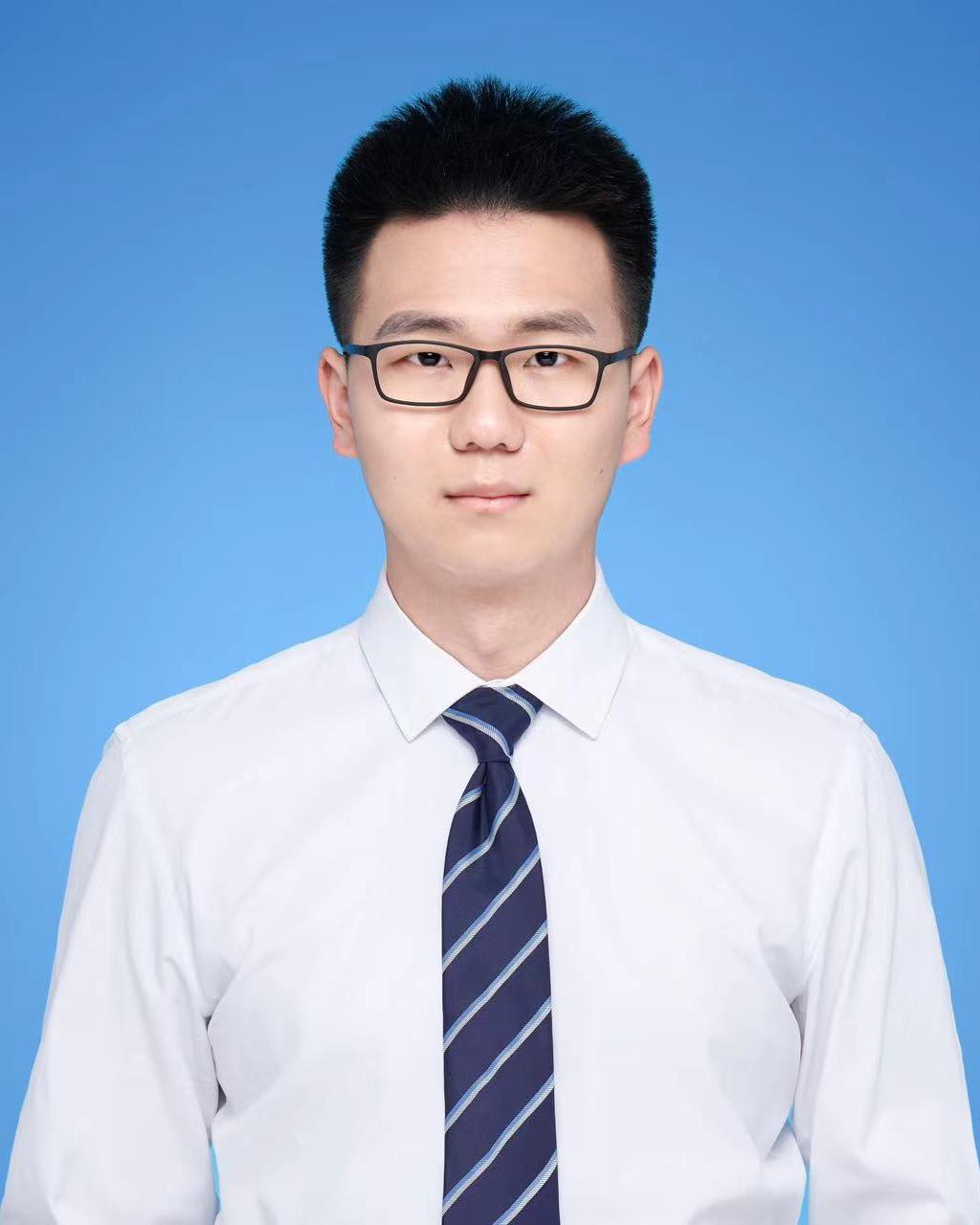}}]{Yiou Wang}

received the B.S. degree in computer science and technology from Beijing University of Technology, Beijing, China, in 2022. He is currently working toward the M.S. degree at the School of Computer Science and Engineering, Beihang University, China. His current research interests include computing-in-memory and deep learning compilers.

\end{IEEEbiography}

\vspace{-8mm}

\begin{IEEEbiography}[{\includegraphics[width=1in,height=1.25in,clip,keepaspectratio]{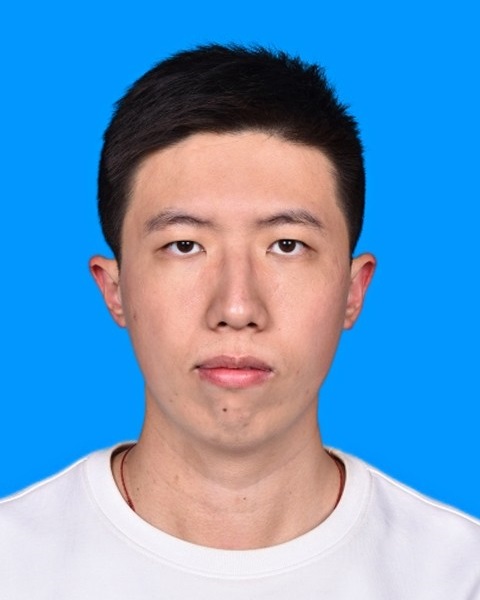}}]{Yingjie Qi}

received the B.S. degree in computer science and technology from Beihang University, Beijing, China, in 2020. He is currently pursuing the Ph.D. degree at the School of Computer Science and Engineering, Beihang University, China. His research interests include graph neural networks acceleration, computing-in-memory architectures and deep learning compilers.

\end{IEEEbiography}

\vspace{-8mm}

\begin{IEEEbiography}[{\includegraphics[width=1in,height=1.25in,clip,keepaspectratio]{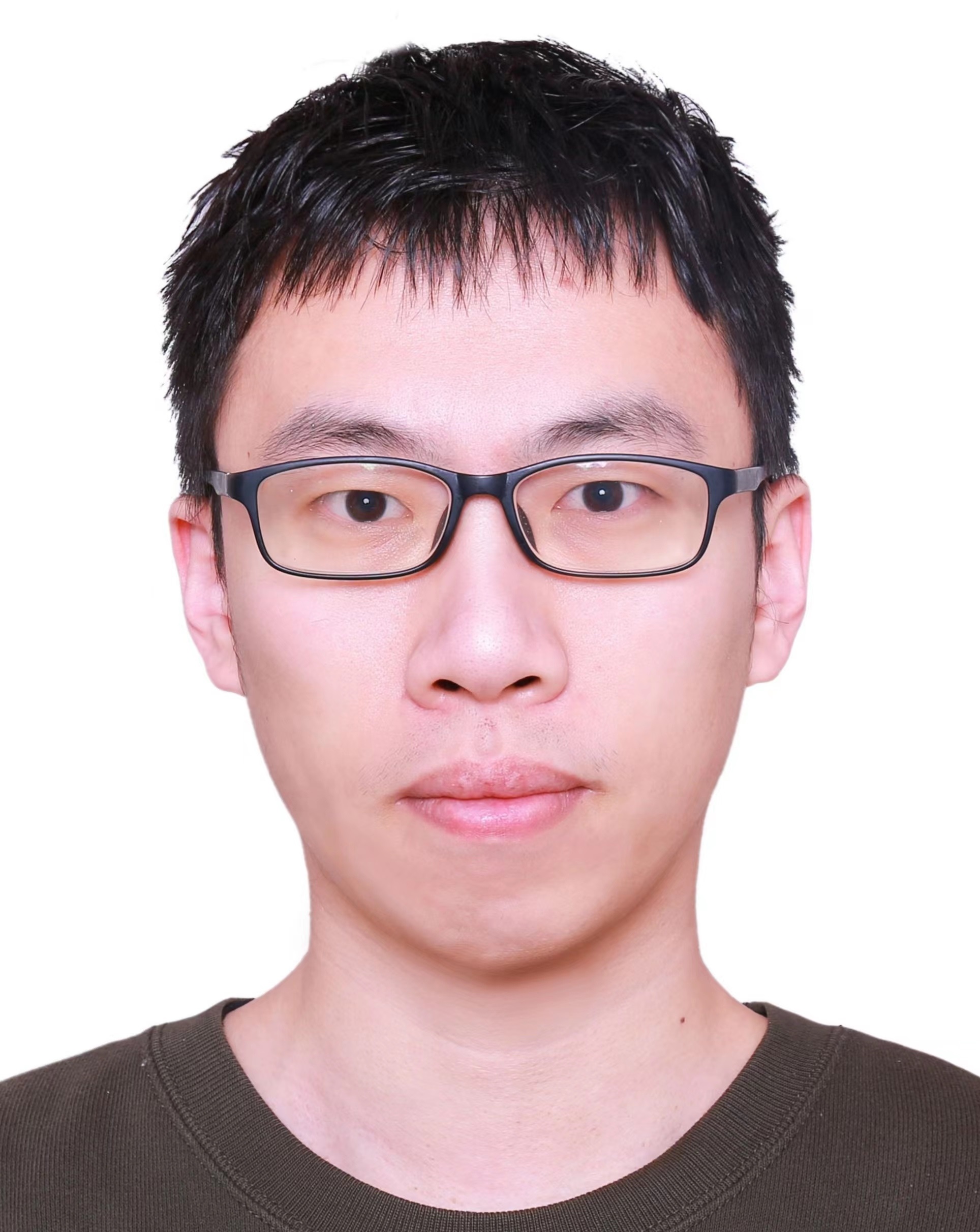}}]{Xiaolin He}

received the B.S. degree in software engineering from Beihang University, Beijing, China, in 2021. He is currently pursuing the Ph.D. degree at the School of Computer Science and Engineering, Beihang University, China. His research interests include processing-in-memory architectures and deep learning compilers.

\end{IEEEbiography}

\vspace{-8mm}

\begin{IEEEbiography}[{\includegraphics[width=1in,height=1.25in,clip,keepaspectratio]{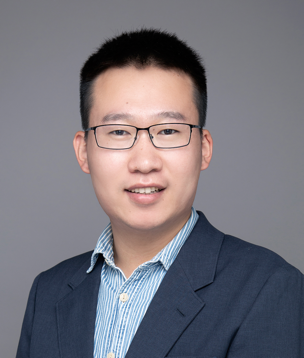}}]{Bonan Yan}

is currently an assistant professor at Institute for Artificial Intelligence, Peking University. He received his PhD degree from Department of Electrical and Computer Engineering, Duke University in 2020. His research interests include circuits and systems for artificial intelligence chips, VLSI design for emerging memory, especially processing-in-memory technology.

He has published more than 40 papers in renowned academic journals and conferences, including ISSCC, Symposium on VLSI Technology, IEDM, DAC, etc. He actively serves as a TPC member for the conferences, including DAC, AICAS, and EDTM.

\end{IEEEbiography}

\vspace{-8mm}

\begin{IEEEbiography}[{\includegraphics[width=1in,height=1.25in,clip,keepaspectratio]{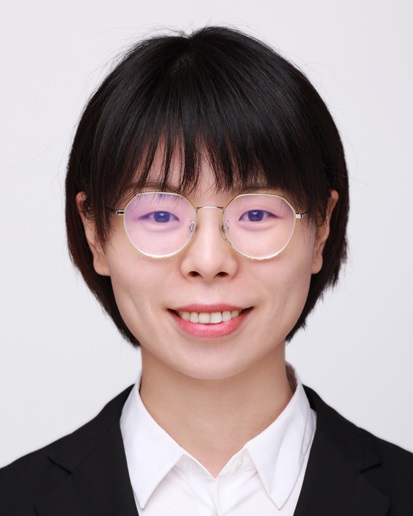}}]{Xueyan Wang}

(S'16-M'19) received the B.S. degree in computer science and technology from Shandong University, Jinan, China,in 2013, and the Ph.D. degree in computer science and technology from Tsinghua University, Beijing, China, in 2018. From 2015 to 2016, she worked as a Research Scholar at the University of Maryland, College Park, MD, USA. She is currently an Associate Professor with the School of Integrated Circuit Science and Engineering, Beihang University. Her primary research interests are in the area of emerging energy-efficient computing architectures, artificial intelligence (AI) chips, and intelligent system. She has authored more than 40 technical papers in leading journals and conferences. She is a recipient of Young Elite Scientists Sponsorship Program by China Association for Science and Technology.

\end{IEEEbiography}

\vspace{-8mm}

\begin{IEEEbiography}[{\includegraphics[width=1in,height=1.25in,clip,keepaspectratio]{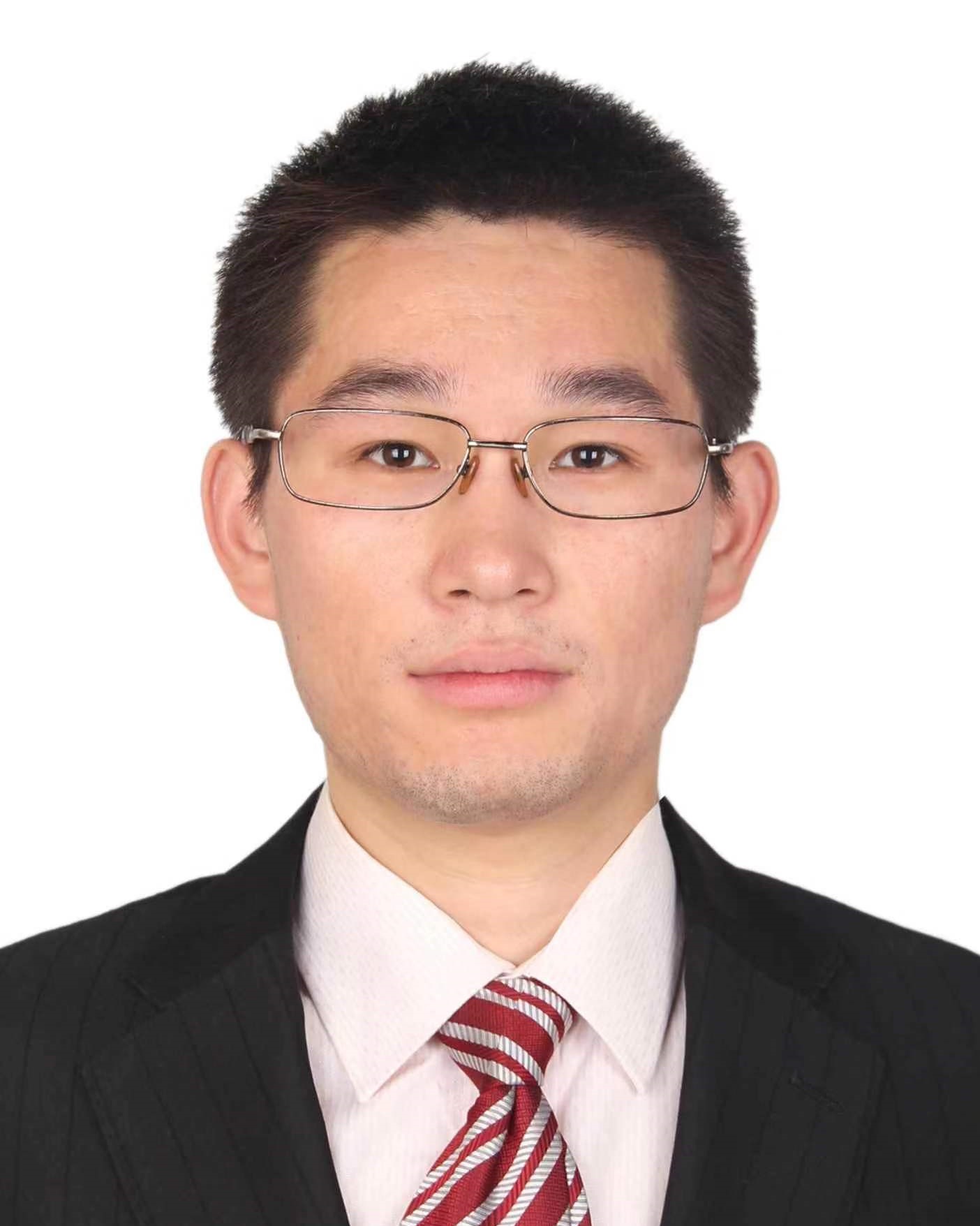}}]{Xiaotao Jia}

(S'13-M'17) received the B.S. degree in mathematics from Beijing Jiao Tong University, Beijing, China, in 2011, and the Ph.D. degree in computer science and technology from Tsinghua University, Beijing, China, in 2016.

He is currently an associate professor in Beihang University, Beijing, China. His current research interests include spintronic circuits and Bayesian learning systems.

\end{IEEEbiography}

\vspace{-8mm}

\begin{IEEEbiography}[{\includegraphics[width=1in,height=1.25in,clip,keepaspectratio]{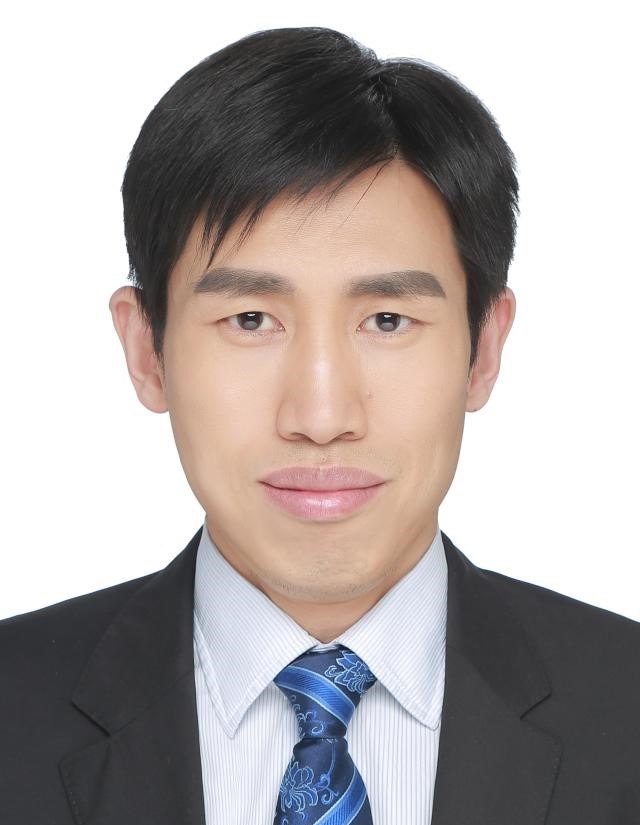}}]{Weisheng Zhao}

(Fellow, IEEE) received the Ph.D. degree in physics from the University of Paris Sud, Paris, France, in 2007.

He is currently a Professor with the School of Integrated Circuit Science and Engineering, Beihang University, Beijing, China. In 2009, he joined the French National Research Center, Paris, as a Tenured Research Scientist. Since 2014, he has been a Distinguished Professor with Beihang University. He has published more than 300 scientific articles in leading journals and conferences, such as \textit{Nature
Electronics}, \textit{Nature Communications}, \textit{Advanced Materials}, IEEE Transactions, ISCA, and DAC. His current research interests include the hybrid integration of nanodevices with CMOS circuits and new nonvolatile memory (40-nm technology node and below), like MRAM circuit and architecture design.

Prof. Zhao was the Editor-in-Chief for the {\sc{IEEE Transactions on Circuits and System I: Regular Paper}} from 2020 to 2023.

\end{IEEEbiography}

\end{document}